\documentclass[useAMS,usenatbib]{mn2e}
\usepackage{epsfig}
\usepackage{epstopdf}
\usepackage{amsmath}
\usepackage{amssymb}
\usepackage{natbib}
\usepackage{lscape}
\newcommand{\yr}{\ensuremath{{\rm yr}}}

\newcommand{\msun}{\ensuremath{{\rm M_{\odot}}}}

\newcommand{\omm}{\Omega_{\rm M}}
\newcommand{\oml}{\Omega_{\rm \Lambda}}

\newcommand{\mfinal}{M_{\rm \star, f}}
\newcommand{\mstar}{M_{\rm \star}}

\newcommand{\sfr}{{\dot{M}_{\rm SFR}}}
\newcommand{\zi}{z_{\rm i}}
\newcommand{\zf}{z_{\rm f}}
\newcommand{\ti}{t_{\rm i}}
\newcommand{\tf}{t_{\rm f}}
\newcommand{\Dt}{\Delta t}
\newcommand{\Dtc}{\Delta t_{\rm c}}
\newcommand{\Dtb}{\Delta t_{\rm b}}
\newcommand{\fml}{f_{\rm ml}}
\newcommand{\fpass}{f_{\rm p}}

\newcommand{\mnow}{M_{\rm \star, 0}}

\newcommand{\lsfrave}{\langle{\log\sfr}\rangle}
\newcommand{\lsfrmed}{\langle{\log\sfr}\rangle_{\rm med}}
\newcommand{\sfrsig}{\sigma_{\rm SFR}}
\newcommand{\ms}{M^{*}}
\newcommand{\Zg}{Z_{\rm g}}
\newcommand{\Zstar}{Z_{\star}}
\newcommand{\Zsun}{Z_{\odot}}

\newcommand{\fsb}{f_{\rm b}}
\newcommand{\Nb}{N_{\rm b}}
\newcommand{\Ppswitch}{P_{\rm p, switch}}
\newcommand{\Psb}{P_{\rm sb}}
\newcommand{\DD}{\Delta_{68}}

\newcommand{\sigX}{\sigma_{\rm X}}
\newcommand{\zhalf}{z_{1/2}}
\newcommand{\zq}{z_{\rm quench}}

\newcommand\ionm[2]{#1{\small\rmfamily{#2}}\relax}% 
\newcommand{\hii}{\ionm{H}{II}}
\newcommand{\nii}{\ionm{N}{II}}
\newcommand{\oiii}{\ionm{O}{III}}
\newcommand{\tlogoh}{12+\log(\mbox{O}/\mbox{H})}
\newcommand{\dd}{{\rm d}}
\usepackage{wasysym}
\usepackage[usenames]{color}
\definecolor{dkgreen}{RGB}{0,200,0}

\begin{document}

\title[Scatter in Galaxy Star Formation Histories]{A Framework for Empirical Galaxy Phenomenology: The Scatter in Galaxy Ages and Stellar Metallicities}

\author[Mu{\~n}oz \& Peeples]{
Joseph~A.~Mu{\~n}oz$^{1, 2}$\thanks{E-mail:jamunoz@physics.ucsb.edu}
and
Molly~S.~Peeples$^{2, 3, 4}$\\
$^1$University of California Santa Barbara, Department of Physics, Santa Barbara, CA 93106, USA\\
$^2$University of California Los Angeles, Department of Physics and Astronomy, Los Angeles, CA 90095, USA\\
$^3$Space Telescope Science Institute; Baltimore, MD 21201, USA\\
$^4$Southern California Center for Galaxy Evolution Fellow\\
}

\maketitle

\begin{abstract}
We develop a theoretical framework that extracts a deeper understanding of galaxy formation from empirically-derived relations among galaxy properties by extending the main-sequence integration method for computing galaxy star formation histories.  We properly account for scatter in the stellar mass--star formation rate relation and the evolving fraction of passive systems and find that the latter effect is almost solely responsible for the age distributions among $z\sim0$ galaxies with stellar masses above $\sim 10^{10}\,\msun$.  However, while we qualitatively agree with the observed median stellar metallicity as a function of stellar mass, we attribute our inability to reproduce the distribution in detail largely to a combination of imperfect gas-phase metallicity and $\alpha$/Fe ratio calibrations.  Our formalism will benefit from new observational constraints and, in turn, improve interpretations of future data by providing self-consistent star formation histories for population synthesis modeling.
\end{abstract}


%-------------------------------------------------------------------------------------------------------------
%             Introduction
%-------------------------------------------------------------------------------------------------------------
\section{Introduction}

The study of galaxy evolution is swimming in a flood of new multi-wavelength data.  Recent observations have characterized (1) the evolution of stellar mass function \citep[e.g.,][]{Bielby12, Muzzin13}; (2) the bimodal nature of galaxies into quiescent and star-forming to $z\sim3$ \citep[e.g.,][]{Brammer11, Muzzin13}; (3) the evolving correlation between the stellar masses and star formation rates of star-forming systems \citep[e.g.,][]{Noeske07, Daddi07, Karim11, Whitaker12, Gonzalez14}; (4) a tight ``fundamental metallicity relation" (FMR) among stellar mass, star formation rate, and gas-phase metallicity \citep[e.g.,][]{KE08, LaraLopez10, Mannucci10, AM13}; (6) the structure of cold gas fueling galaxies \citep[e.g.,][]{Genel10}; and (6) the environmental influence on galactic properties \citep[e.g.,][]{Peng10, Pasquali10, Lin14}.  Yet, these various empirical relations beg a theoretical framework that answer questions about their effects on and relative importance in the buildup of stellar mass in the universe.  For example, what role does the quiescent phase of galaxy evolution play in setting the growth of the stellar population, and do clumps in cold streams or environmental processes fundamentally drive star formation?

Moreover, understanding the history of star formation in the universe is critical, not only for its own sake, but because the knowledge feeds back into the interpretation of the torrent of extragalactic observations.  Extracting most galaxy properties from the data, for example, requires stellar population synthesis models for which the star formation history is a necessary input \citep[e.g.,][]{Tinsley80,Leitherer99, Schaerer13}.  These histories are also central to understanding the nature of the mass-metallicity relation \citep[e.g.,][]{PS13} and the evolution of the law relating star formation rate to gas density \citep[e.g.,][]{Bigiel08}. 

To avoid the degeneracies and sub-grid prescriptions involved in formulating a completely theoretical model of galaxy formation and evolution for comparison with observations, we are motivated to organize the empirical data itself into a cohesive structure describing the buildup of stellar mass within galaxies.  Such a path is provided by the Main Sequence Integration (MSI) method \citep[e.g.][]{Leitner12}.  This scheme traces the star formation history of a mock galaxy by using the observed average star formation rate as a function of stellar mass and redshift to inform the growth of the system in each time step.  It can also track the buildup of stellar metallicity by incorporating the empirically-determined FMR \citep{PS13}.  However, the current method ignores the significant scatter in these empirical relations \citep[e.g.,][]{Noeske07, Daddi07, Karim11, Guo13, Gonzalez14} as well as the quenching of star formation implied by the evolving fraction of quiescent galaxies.  Both effects contribute to large variations in the star formation history from system-to-system, which in turn, imprint on the derivations of galaxy properties from the data.  

To study the influence of empirical phenomenology on the star formation process more fully, we develop an improved MSI model that introduces fluctuations into the star formation process and, by construction, reproduces both the scatter in the stellar mass--star formation rate relation and the evolving fraction of quiescent galaxies.  In \S\ref{sec:basic}, we review the basic MSI method, which traces the mean history of smoothly and continuously star-forming systems.  We present our new prescriptions for generating variations in the star formation history, both in time and from galaxy-to-galaxy, in \S\ref{sec:scatt}, and test our method using Sloan Digital Sky Survey (SDSS) observations from \citep{Gallazzi05} of the scatter in galaxy ages and stellar metallicities in \S\ref{sec:results}.  Finally, we discuss and summarize our conclusions in \S\ref{sec:forward}.

Throughout, we adopt a $\Lambda$CDM cosmology with $\omm=0.28$, $\oml=0.72$, and $H_0=70\,{\rm km/s/Mpc}$ and a \citet{Chabrier03} initial mass function (IMF).  Furthermore, $\log$ always refers to the base-10 logarithm.

%-------------------------------------------------------------------------------------------------------------
%             The Basic Model
%-------------------------------------------------------------------------------------------------------------
\section{The Basic Model}\label{sec:basic}

We begin by reviewing the basic formulation of the MSI model.  In \S\ref{sec:basic:sfh}, we describe the computation of the average star formation history of a galaxy with final stellar mass $\mfinal$ at redshift $\zf$ assuming no scatter in the star formation rate for a given stellar mass and redshift.  Then, in \S\ref{sec:basic:mh}, we discuss the metallicity history of galaxies using the fundamental metallicity relation between stellar mass, star formation rate, and gas-phase metallicity.

%-----------Star Formation Histories----------------------------------------------------------------------------------
\subsection{Star Formation Histories}\label{sec:basic:sfh}

In the MSI procedure \citep[e.g.,][]{Leitner12}, the observed relationship between star formation rate, $\sfr$, and stellar mass, $\mstar$, as a function of redshift determines galactic star formation histories.  Explicitly, in the time $\dd t$ between redshifts $z+\dd z$ and $z$, the stellar mass, $\mstar$, of a galaxy increases by an amount $\sfr(\mstar,z)\,\dd t$, where $\dd t$ is the interval of cosmic lookback time, $t$, corresponding to the redshift interval $\dd z$ given by
\begin{equation}\label{eq:time}
\dd t = \frac{\dd z}{H_{0}\,(1+z)\,\sqrt{\omm\,(1+z)^3+\oml}}.
\end{equation}
Using equation~\ref{eq:time} and our chosen cosmology we will refer to $t$ and $z$ interchangeably throughout this work.

We use the $\sfr$--$\mstar$ relation recently derived by \citet{Whitaker12} from galaxies in the NEWFIRM Medium-Band Survey \citep{Whitaker11}.  These authors compute star formation rates from a combination of the rest-frame UV and IR emission and stellar masses assuming a \citet{Chabrier03} IMF, solar metallicity, and exponentially declining star formation histories.\footnote{Note that the star formation and enrichment histories assumed by the analysis in \citet{Whitaker11} will be inconsistent with our results.}  The resulting relation, which is similar to that derived using a comparable analysis of galaxies in the COSMOS survey by \citet{Karim11}, has a median logarithmic star formation rate for star-forming galaxies well-fit by
\begin{equation}\label{eq:sfr}
\lsfrmed=\alpha(z)\,\left[\log(\mstar/\msun) - 10.5\right]+\beta(z),
\end{equation}
where $\alpha=0.70-0.13\,z$ and $\beta=0.38+1.14\,z-0.19\,z^2$.  Despite the limited mass range over which this formula was derived, particularly at the highest redshifts, we extrapolate equation~\ref{eq:sfr} in the present work to any combination of redshift and stellar mass required.  With this choice and in the MSI formulation without scatter, the mass added in each time step $\dd t$ at redshift $z$ is $10^{\lsfrmed(\mstar,z)}\,\dd t$.\footnote{Because the scatter about the median logarithmic star formation rate is roughly symmetric, we ignore the small difference between the median and average.}

In addition to the mass gained in each time step, \citet{LK11} considers the mass recycled back into the interstellar medium as a result of stellar winds and death.  We quantify the amount of mass lost in this way by the fraction of stellar mass, $\fml(t_{2}-t_{1})$, produced at time $t_{1}$ that has been lost by some later time $t_{2}$, and adopt the fit to this fraction which \citet{LK11} derive using population synthesis modeling for a \citet{Chabrier03} IMF at solar metallicity:
\begin{equation}\label{eq:fml}
\fml=C_{0}\,{\rm ln}\left(\frac{t_{2}-t_{1}}{\lambda}+1\right),
\end{equation}
where $C_{0}=0.046$ and $\lambda=2.76\times 10^5\,{\yr}$.  
The dominant source of uncertainty in equation~\ref{eq:fml} is the choice of IMF since the total mass loss, unlike the amount lost in each individual stage of stellar evolution, is relatively insensitive to metallicity.
Given this fitting, the instantaneous rate of mass recycling at time $t$ is a convolution of $\fml$ and the star formation history $\sfr(t)$.  More explicitly, for a star formation history beginning at some time $\ti$, the total mass loss rate at time $t>\ti$ from stars formed between $\ti$ and $t$ is 
\begin{equation}\label{eq:mrec}
\frac{\dd M_{\rm rec}}{\dd t}=\int_{\ti}^{t}\!\! \sfr(t')\,\frac{\dd\fml(t-t')}{\dd t'}\,\dd t'.
\end{equation}
Thus, the change in stellar mass between from $z+\dd z$ to $z$ now becomes $\left[10^{\lsfrmed(\mstar,z)}-\dd M_{\rm rec}(z)/\dd t\right]\,\dd t$.

In this work, we are more interested in the origins of observed galaxies today rather than the eventual fates of high-redshift galaxies.  Therefore, we follow \citet{LK11} in tracing star formation histories backward rather than forward in time and beginning at $z=0$.  However, this procedure complicates the evaluation of equation~\ref{eq:mrec} at late times since it requires foreknowledge of the star formation history at early times.  \citet{LK11} circumvent this problem using an iterative approach to solve for the star formation history, and we adopt the same solution.  We initially assume $\fml=0$ and compute the star formation history without mass recycling.  We then use this history to reevaluate equation~\ref{eq:mrec} and trace the star formation history in the subsequent iteration.  This process typically converges quickly, after only a few iterations.  An example history for a galaxy that has $\mstar=10^{10.5}\,\msun$ at $z=0$ is shown as the thick black line in Figure~\ref{fig:sfh}.

We calculate the star formation history of a galaxy beginning our calculations at time $t=\tf$, corresponding to $z=\zf$, and take steps backwards in time until we reach some fiducial ``initial'' time $\ti$ at which our galaxies are ``seeded'' with some relatively small fiducial stellar mass with zero age and uniform metallicity.  We compute the galactic average age, $T$, of the stellar population at $\tf$ by taking a mass-weighted average of the mass surviving from each time step, i.e.,
\begin{equation}\label{eq:age}
T=\int_{\ti}^{\tf}\!\! \frac{\sfr(t)}{\mstar(\tf)}\,[1-\fml(\tf-t)]\,(\tf-t)\,\dd t.
\end{equation}
The implicit assumption that stars already present in a galaxy at $\ti$ all formed at time $\ti$ does not significantly affect the results if $\zi$, the redshift corresponding to $\ti$, is large and the stellar mass at $\zi$ small (see Appendix~\ref{sec:app:res}).  Typically, $\mstar(\ti)/\mstar(\tf)<10^{-3}$ in our realizations. 

In the interest of making out results more comparable to observations, we can also calculate galaxy-averaged ages weighted by {\it{B}}-band luminosity rather than by mass using a model for the stellar metallicities within a mock galaxy (see \S\ref{sec:basic:mh}).\footnote{Note that this is not precisely equivalent to the methodology of \citet{Gallazzi05}, who derive stellar ages, metallicities, and masses from a set of spectral indices (see \S\ref{sec:results}).}  To do this, we replace the fraction of final mass produced in each a step $\dd t$---$\sfr(t)\,[1-\fml(\tf-t)]\,\dd t/\mstar(\tf)$---which appears in each of equations~\ref{eq:age}, with the fraction of the total {\em B}-band luminosity emitted by these stars at time $\tf$.  As in \citet{PS13}, we compute this luminosity for stars in each time step given its age and metallicity using \citet{BC03} stellar population models.  The resulting luminosity weighting significantly biases the inferred properties toward more recent star formation and, hence, produces younger ages \citep{TS09} and higher metallicities.

%-----------Metallicity Histories----------------------------------------------------------------------------------
\subsection{Metallicity Histories}\label{sec:basic:mh}

In addition to the average age, we also calculate the average metallicity, $\Zstar$, of a galaxy's stellar population from its star formation history.  Following \citet{PS13}, we assume that the metallicity of the galactic gas, $\Zg$, in a given time step imprints on the stars that form out of it at that time step \citep[see also][]{Tinsley75}.  This procedure assumes, similar to our calculation of the average age, that all stars already present at time $\ti$ have the same metallicity given by the gas-phase metallicity at that initial time, but this does not significantly affect the results for large $\zi$ and small $\mstar(\zi)$.  At $\tf$, the mass-weighted, galaxy-averaged stellar metallicity is
\begin{equation}\label{eq:OHave}
\Zstar=\int_{\ti}^{\tf}\!\! \frac{\sfr(t)}{\mstar(\tf)}\,[1-\fml(\tf-t)]\,\Zg[\mstar(t),\sfr(t)]\,\dd t.
\end{equation}
That is, stellar metallicities are the star formation-weighted histories of gas metallicities \citep{PS13}.  We can also compute a luminosity-weighted version of $\Zstar$ using the same procedure as for $T$.

We specify $\Zg$ as a function of both $\mstar$ and $\sfr$ using the so-called ``fundamental metallicity relation'' (hereafter FMR) observed at $z=0$, which connects stellar mass, star formation rate, and gas-phase metallicity \citep[e.g.,][]{Mannucci10, LaraLopez10}.  By incorporating the influence of star formation on metallicity, this choice represents an improvement over assigning metallicities based on an evolving relationship with stellar mass alone \citep[e.g.,][]{Zahid12b}.  As in \citet{PS13}, we assume that the FMR is redshift-independent, though we emphasize that the current observational situation is unclear \citep[e.g.,][]{Yuan13a, Zahid13, Yabe14, dlReyes14}.  

Despite a theoretical motivation for the relation \citep[e.g.][]{Forbes14}, the formulation of the FMR is purely empirical and, thus, complicated by the physics of observational metallicity indicators.  Recently, \citet{AM13} showed that the choice of calibrator can strongly influence the dependence of the inferred metallicity on not only stellar mass \citep{KE08}, but also on the star formation rate.  This effect is due to the sensitivity of metal lines to the electron temperature and excitation/ionization structure of \hii\ regions, which in turn depend on $\sfr$ \citep[e.g.,][]{Pena12, AM13}.  Moreover, incorrect parameterization of these properties can introduce systematic uncertainties into the derived metallicity measurement, though \citet{Nicholls12} and \citet{Dopita13} show that adding a high-energy tail to the standard Maxwell--Boltzmann distribution of electron energies can improve temperature estimates and reconcile seemingly discrepant indicators.

However, as the FMR has not yet been fully characterized using this new method, we instead investigate how our results vary when adopting several different representative calibrations whose dependence on $\mstar$ and $\sfr$ span the range described in \citet{AM13}.  Specifically, we take
\begin{equation}\label{eq:fmr}
\tlogoh = \beta (\log\mstar - \alpha\sfr) + \gamma
\end{equation}
and set the gas-phase metallicity relative to solar, $\Zg/\Zsun$, to be the relative oxygen-to-hydrogen ratio, (O/H)/(O/H)$_{\odot}$, with $12+\log(\rm O/H)_{\odot}=8.76$ \citep{Caffau11}.  \citet{AM13} find $(\alpha, \beta, \gamma)=(0.325, 0.308, 5.562)$ using the [\oiii]--[\nii] calibration of \citet[][hereafter PP04]{PP04}; $(0.125, 0.208, 6.827)$ for the calibration of \citet[][hereafter KD02]{KD02}, which uses the upper branch of the $R_{23}$ diagnostic; and $(0.248, 0.358, 5.363)$ for the calibration of \citet[][hereafter Z94]{Zaritsky94}.  These three calibrations span the uncertainties in both the normalization of the FMR and its dependence on star formation rate with the PP04 fitting producing an approximately average relationship.  Moreover, by stacking spectra from SDSS, \citet{AM13} are able to characterize the FMR down to stellar masses nearly 2~dex lower than that in earlier studies.  Adopting their relations, we are less susceptible to extrapolation errors than were \citet{PS13}.  Therefore, where not otherwise specified in \S\ref{sec:results}, we assume the FMR as measured by \citet{AM13} and choose the [\oiii]--[\nii] calibration of PP04 as fiducial.

We also consider results using the double quadratic fit to the FMR introduced by \citet[][hereafter M10]{Mannucci10}: 
\begin{equation}\label{eq:fmrM10}
\begin{split}
\tlogoh=&8.90+0.37\,m_{10}-0.19\,m_{10}^2 \\ 
&-0.14\,\dot{m}_{1}-0.054\,\dot{m}_{1}^2+0.12\,m_{10}\,\dot{m}_{1},
\end{split}
\end{equation}
where $m_{10}=\log(\mstar/10^{10}\,\msun)$ and $\dot{m}_{1}=\log(\sfr/\msun\,\yr^{-1})$.  By assuming that star-forming galaxies always remain on this relation, \citet{PS13} are able to approximately reproduce the $z=0$ stellar metallicities of $10^{10.5}\msun$ galaxies observed by \citet{Gallazzi05} using the average MSI-derived star formation history of \citet{Leitner12}.  However, as we show in \S\ref{sec:results:metals}, this agreement may have been partially coincidental since a significant fraction of galaxies in this mass range are passive and, thus, have substantially different star formation histories (and corresponding enrichment histories) than \citet{PS13} assumed.  These authors further attribute their over-estimate of the stellar metallicities at $\lesssim 10^{10}\msun$ to extrapolation errors in both the star formation histories and the M10 FMR and interpreted their results as producing good agreement with observations.  

However, we must also be aware that the gas-phase metal abundances of the FMR are measured relative to oxygen, an $\alpha$ element.  Stellar metallicities, on the other hand, are typically measured from iron lines.  Because $\alpha$ elements are almost exclusively produced in Type~II supernovae while a significant fraction of iron is additionally made in Type~Ia supernovae, the two metallicities are not directly comparable or even related by a constant factor.  Observational evidence suggests that the delayed contribution of Type~Ia supernovae to the iron budget produces super-solar $\alpha$/Fe ratios in older, more massive galaxies \citep{Thomas05, Arrigoni10}, but \citet{Stoll13} purport an intrinsic correlation between $\alpha$/Fe and the oxygen abundance, O/H, based on measurements of individual halo, bulge, and disk stars in the Milky Way.  We will use the empirical fit from \citet{Stoll13}, given by
\begin{equation}\label{eq:alpha_Fe}
\log\frac{\rm (Fe/H)}{\rm (Fe/H)_{\odot}}=-0.34+1.25\,\log\frac{\rm (O/H)}{\rm (O/H)_{\odot}},
\end{equation}
for our results in \S\ref{sec:results} as a fiducial case but also discuss the effect of a mass-dependent conversion.

%-------------------------------------------------------------------------------------------------------------
%             Incorporating Scatter
%-------------------------------------------------------------------------------------------------------------
\section{Incorporating Scatter}\label{sec:scatt}

Going beyond the previous work by \citet{Leitner12} and \citet{PS13}, in this study, we consider the scatter in star formation histories from galaxy-to-galaxy indicated both by variations in the star formation rate at a given stellar mass and redshift and by the evolving fraction of quiescent systems.  In this section, we describe the implementation of these fluctuations in our version of the MSI model.  In \S\ref{sec:results}, we will test whether and by what mechanisms the resulting scatter in the star formation history imprints on the present-day mass-age and mass-metallicity relations measured by \citep{Gallazzi05}.  Because of this, we will generally set $\zf=0$ and denote $\mnow \equiv \mstar(\zf=0)$ unless otherwise specified.

At each time step of our MSI formulation, instead of specifying the star formation rate via equation~\ref{eq:sfr}, we draw a value from a probability distribution that may depend on stellar mass and/or redshift.\footnote{Incorporating scatter into our calculations of galaxy star formation histories requires generating many random numbers.  However, as outlined in \S\ref{sec:basic:sfh}, our procedure also involves iteration to determine consistently the amount of stellar mass recycled back into the ISM.  To ensure convergence, note that we must preselect our sets of random numbers that specify the specific star formation history of a galaxy before the first iteration and maintain the same set throughout the convergence process.}  In this section, we describe how we piece together this distribution from components of varying sophistication and physical motivation.  Note that this probability distribution may not have a mean given by equation~\ref{eq:sfr} since the observations to which we compare our final results need not be completely described by only normal, star-forming galaxies.  Moreover, successive draws from this distribution may be correlated to simulate different amounts of stochasticity.

\begin{figure}
\begin{center}
\includegraphics[width=\columnwidth,trim=30 20 235 0,clip]{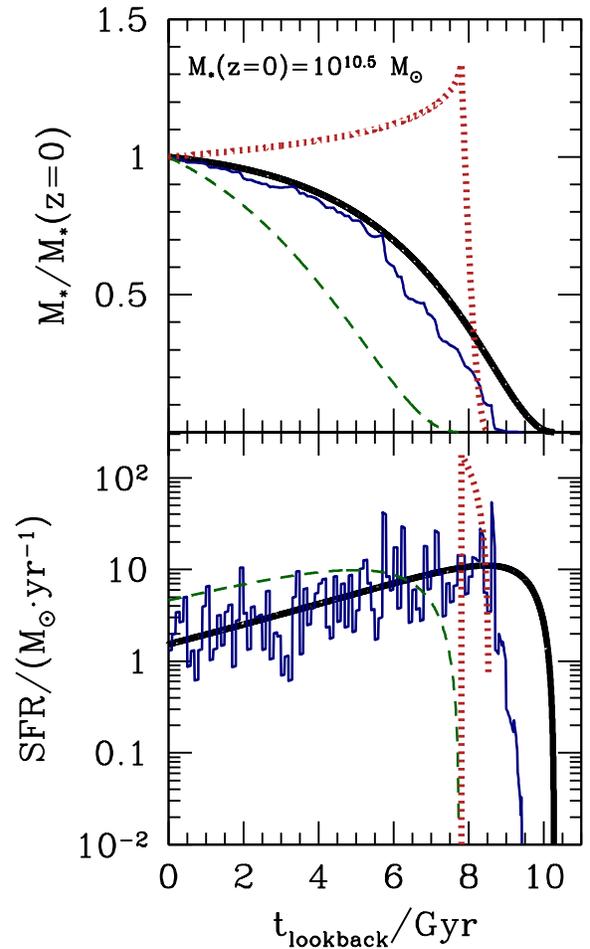} 
\caption{\label{fig:sfh} 
Example model star formation histories showing the stellar mass (upper panel) and star formation rate (lower panel) as a function of lookback time.  We set the final stellar mass for these examples to be $10^{10.5}\,\msun$ at $z=0$.  The thick, solid (black) curve shows the mean star formation history computed using only the average star formation rate as a function of stellar mass and redshift.  The thin solid (blue) and short-dashed (green) curves include 0.3 dex of scatter in the star formation rate correlated over a time-scale of, respectively, $\Dtc=10^{8}\,\yr$ and $10^{10}\,\yr$ (effectively its entire lifetime).  The dotted (red) curve is an example with $\Dtc=10^{9}\,\yr$ and passive galaxies included; this particular example becomes passive quite early and slowly loses stellar mass from stellar death and recycling over the last 8 Gyr.
}
\end{center}
\end{figure}

Figure~\ref{fig:sfh} shows example star formation histories for galaxies with a $z=0$ stellar mass of $\mnow=10^{10.5}\,\msun$, about that of the Milky Way \citep[e.g.,][]{Klypin02}. The solid, black curve demonstrates the mean relation, incorporating no scatter in the star formation rate about the mean determined from equation~\ref{eq:sfr}.  The other modeled star formation histories in Figure~\ref{fig:sfh} account for the scatter in star formation rates for normal, star forming galaxies (\S\ref{sec:scatt:norm}) and the evolving population of passive galaxies (\S\ref{sec:scatt:passive}).

%-------------------------------------------------------------------------------------------------------------
\subsection{Normal, Star-Forming Galaxies}\label{sec:scatt:norm}

Observationally, the star formation rate of star-forming galaxies at fixed stellar mass has an approximately log-normal distribution with a logarithmic standard deviation, $\sfrsig$, of approximately 0.3 dex, roughly independent of stellar mass and redshift \citep[][though see \citealp{Guo13}]{Noeske07, Daddi07, Karim11, Gonzalez11, ML11, Munoz12}.  Unfortunately, the tails of the distribution are not well-characterized observationally, and the amount of scatter may be as high as 0.4 dex or as low as 0.2.  However, we find that our results are relatively insensitive to this uncertainty (see Appendix~\ref{sec:app:test}), and, thus, assume a fiducial value of $\sfrsig=0.3$.  Assuming such a shape, we take the mean of the distribution to be identical to the observed median given in equation~\ref{eq:sfr}.  

Tracing the star formation history with an integration time step $\Dt$ (see Appendix~\ref{sec:app:res}), we resample the star formation rate from our log-normal distribution every $\Dtc/\Dt$ time steps by selecting the size of the fluctuation, $(\log \sfr- \lsfrave)/\sfrsig$.  In this way, a particular fluctuation affects the galaxy for a time $\Dtc$, and we allow the star formation rate itself to change on scales of $\Dt$ as $\lsfrave$ evolves with redshift and the growing mass of the galaxy.  The time-scale $\Dtc$, thus, is a critical parameter of our model and controls how correlated the star formation rate is through time.  For a large value, a particular fluctuation in the star formation rate about the mean will persist for a more significant period and more significantly affect the final properties.  On the other hand, a small $\Dtc$ results in less scatter in the final properties, since an aberrant star formation rate in one time step is allowed to continue for a shorter period before being resampled.  Though we are not testing a physical model here, different correlation time-scales may have distinct physical geneses; environmental effects likely operate over long time-scales, for example, while instabilities in the inflow rate of cold gas correlate star formation over short time-scales.

In Figure~\ref{fig:sfh}, we show two model examples of star formation histories that include only normal star formation.  The thin solid (blue) and short-dashed (green) curves assume $\Dtc=10^{8}$ and $10^{10}\,\yr$, respectively.  For $\Dtc=10^{8}\,\yr$, the star formation history does not deviate much from the ``no scatter'' case since strong fluctuations in star formation rate are not allowed to persist for a significant amount of time.  On the other hand, for $\Dtc=10^{10}\,\yr$ (effectively the entire galaxy lifetime) the history is rather smooth because the distribution is only sampled once ($\log \sfr - \lsfrave\sim1.5\,\sfrsig$ for the example in Fig.~\ref{fig:sfh}).  A high values of $\Dtc$ with a single fluctuation per system generates much stronger deviations from galaxy to galaxy than does a low value, which combines many uncorrelated fluctuations.  However, we emphasize that, by construction, an ensemble of our output star formation histories reproduce the observed stellar mass--star formation rate relation and its scatter for any choice of $\Dtc$.  In \S\ref{sec:results}, we compare results for several different values, which may reflect different physical mechanisms driving the scatter in the star formation rate.

%-------------------------------------------------------------------------------------------------------------
\subsection{Passive Galaxies}\label{sec:scatt:passive}

While previous implementations of the MSI method considered only star-forming galaxies, the \citet{Gallazzi05} data to which we compare our results in \S\ref{sec:results} includes passive galaxies.  The ``quenching" of star formation can have a strong influence on the scatter among the histories of massive galaxies by suppressing galaxy star formation rates at different times.  We construct a new prescription to incorporate this suppression into the MSI method that allows us to quantify the resulting effect on stellar ages and metallicities.  

As with star-forming galaxies, we calibrate our model based on empirical observations from the NEWFIRM Medium-Band Survey, using the passive galaxy fraction measured by \citet{Brammer11} at $0.4\leq z\leq 2.2$ for stellar masses above $\sim10^{9.5}$.  These data were recalibrated and fit by \citet{Behroozi13}, who found that the fraction of galaxies that are passive depends on stellar mass and redshift as
\begin{equation}\label{eq:fpass}
\fpass(\mstar,z)=\left[\left(\frac{\mstar}{10^{10.2+0.5\,z}\,\msun}\right)^{-1.3}+1\right]^{-1}.
\end{equation}
Because equation~\ref{eq:fpass} implies that virtually all dwarf galaxies are star-forming, it may not adequately describe the quenching of star formation in local group satellites as a result of ram-pressure stripping \citep[e.g.,][]{Kravtsov04, Munoz09}.  Nevertheless, we extrapolate this fitting to all necessary masses and redshifts in our model.  Moreover, we compute results specifically for star-forming populations in \S\ref{sec:results} by setting $\fpass(\mstar,z)=0$.  

However, the passive fraction is only the desired result of our prescription; we cannot simply use $\fpass$ to select quenched systems at each time step since this would allow them to alternate between passive and star-forming phases on the scale of $\Dt$.  Instead, in our scheme, we assume that galaxies exhibit a ``once passive, always passive'' behavior.  (Since we integrate star formation histories backward through time, this is equivalent to a rule of ``once star-forming, always star-forming.")  Moreover, galaxies designated as passive form no new stars but still lose stellar mass due to stellar recycling according to equation~\ref{eq:mrec}.  The procedure is as follows.  Beginning at $z=0$, we determine whether a galaxy is passive or not based on the probability given by equation~\ref{eq:fpass} as a function of the galaxy's mass, $\mnow$.  A galaxy marked as star-forming at $z=0$ is then star-forming for its entire history.  On the other hand, for a galaxy set to be passive at $z=0$, we calculate the probability in each time step that the system switches from passive to star-forming based on its stellar mass and the redshift of the time step.  While $\fpass$ gives the fraction of all galaxies that are passive, the probability that a passive galaxy becomes star-forming---again as we integrate backward through time---in a redshift interval $\dd z$, is given by the change in the number of passive galaxies over $\dd z$ as a fraction of the total number of passive galaxies, which can be written as 
\begin{equation}\label{eq:Ppswitch}
\Ppswitch(\mstar, z)=-\frac{\dd}{\dd z}\ln\left(\fpass\,\frac{\dd n}{\dd\mstar}\right)\,\dd z,
\end{equation}
where $\dd n/\dd\mstar$ is the stellar mass function.  Equation~\ref{eq:Ppswitch} implicitly includes the contribution to the changing number of passive galaxies owing to the evolution in the total number of galaxies.  Once a passive galaxy becomes star-forming, its star formation rate is then selected using the previously described procedure (\S\ref{sec:scatt:norm}).  This prescription ensures that, averaged over a large and representatively-sampled population of model galaxies, the fraction of systems denoted as passive will be given by equation~\ref{eq:fpass}.   

All that remains to describe our implementation of passive galaxies completely is to specify our choice of $\dd n/\dd\mstar$.  We adopt the observed stellar mass functions of \citet{Bielby12}, who use data from the WIRCam Deep Survey and a methodology for deriving stellar masses very similar to that of \citet{Brammer11} and \citet{Whitaker12} to fit results at discrete redshifts from $0.2< z < 2$ with a double-\citet{Schechter76} function:
\begin{equation}\label{eq:smf}
\frac{\dd n(\mstar)}{\dd\ln\mstar} = {\rm e}^{\mstar/\ms}\,\left[\phi_{1}\,\left(\frac{\mstar}{\ms}\right)^{\alpha_{1}}+\phi_{2}\,\left(\frac{\mstar}{\ms}\right)^{\alpha_{2}}\right]\,\left(\frac{\mstar}{\ms}\right).
\end{equation}
To obtain a continuous interpolation at any value of $z$, we fit the redshift evolution of the five free parameters in equation~\ref{eq:smf} as
\begin{equation}\label{eq:smf1}
\log(\ms/\msun) = -0.114\,z+10.79, \nonumber
\end{equation}
\begin{equation}
\phi_{1} = (-0.96\,z+2.76)\times10^{-3}, \nonumber
\end{equation}
\begin{equation}
\phi_{2} = 0.23\times10^{-3}, \nonumber
\end{equation}
\begin{equation}
\alpha_{1} = 0.66\,z-1, \nonumber
\end{equation}
and
\begin{equation}\label{eq:smfparam}
\alpha_{2} = \begin{cases} -0.5\,z-1.5, & z\le1 \\ -2, & z>1 \end{cases}.
\end{equation}
Equations~\ref{eq:smf} and~\ref{eq:smf1} give a stellar mass function consistent with those from other recent works \citep[e.g.,][]{Brammer11, Muzzin13}.  Because we only require the redshift evolution of the mass function in our model via equation~\ref{eq:Ppswitch}, our specific choice---in addition to effects such as cosmic variance---is even less significant to our results.

Figure~\ref{fig:sfh} shows two examples (both with $\Dtc=10^{9}\,\yr$) of star formation histories where we have included a passive phase.  In one case (dotted, red), the galaxy becomes massive very early in its lifetime and then evolves only quiescently for nearly 8 Gyr until $z=0$.  Once star formation ceases, the stellar mass declines somewhat due to stellar recycling before reaching $\mnow=10^{10.5}\,\msun$.  By contrast, in the other example (long-dashed, orange), the galaxy builds up its mass much more gradually and only becomes passive relatively recently in cosmic time---less than 500 Myr ago.  As we show in \S\ref{sec:results}, this variation adds a significant amount of scatter to the ages and stellar metallicities of galaxies at this stellar mass and increases the median age of systems more massive than $\sim 10^{10}\,\msun$.

%-------------------------------------------------------------------------------------------------------------
%             Results
%-------------------------------------------------------------------------------------------------------------

\begin{figure*}
\begin{center}
\includegraphics[width=\textwidth,trim=25 200 30 0,clip]{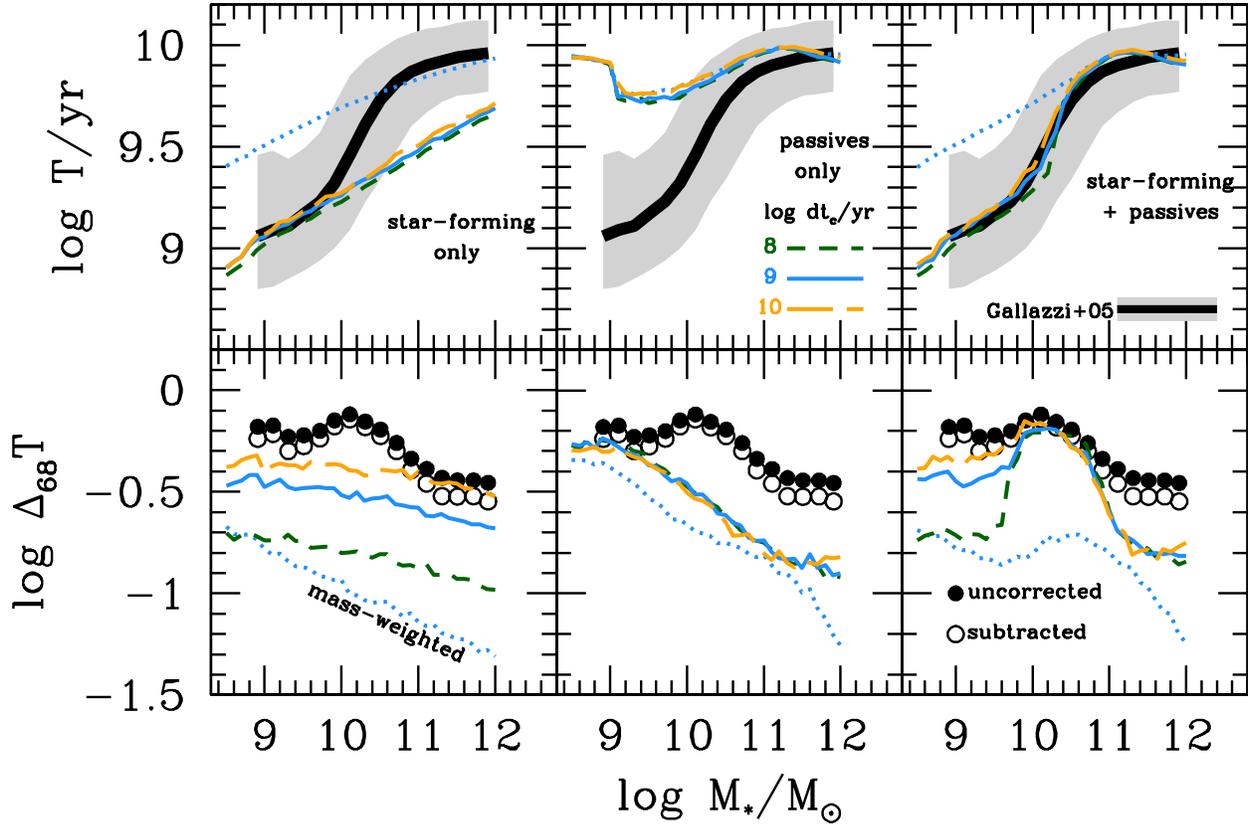}
\caption{\label{fig:scatter1} 
The median (top row) and central 68\% interval range (bottom row) for galaxy-averaged stellar age as a function of total stellar mass at $z=0$.  Thick lines and shaded regions show the observations by \citet{Gallazzi05} for a representative sample of both star-forming and passive galaxies in the top row, while filled and open circles in the bottom row indicate the central 68\% interval uncorrected and uncertainty-subtracted via Eq.~\ref{eq:DXcorr}, respectively.  Model results are based on $10^{3}$ realizations of star formation histories per stellar mass bin.  Short-dashed (green), solid (blue), and long-dashed (orange) curves compare the effect of different correlation time-scales between fluctuations in normal star formation: $\Dtc=10^{8}$, $10^{9}$, and $10^{10}\,\yr$, respectively.  Dotted (blue) curves indicate results for a mass-weighted averaging of the stellar ages within a galaxy with $\Dtc=10^{9}\,\yr$, while all other lines show luminosity-weighted results.  Model results include only star-forming galaxies in the left column and only galaxies that are quenched by $z=0$ in the middle column, while results in the right column reflect a representative population with both star-forming and passives systems.
}
\end{center}
\end{figure*}

\begin{figure*}
\begin{center}
\includegraphics[width=\textwidth,trim=30 40 20 0,clip]{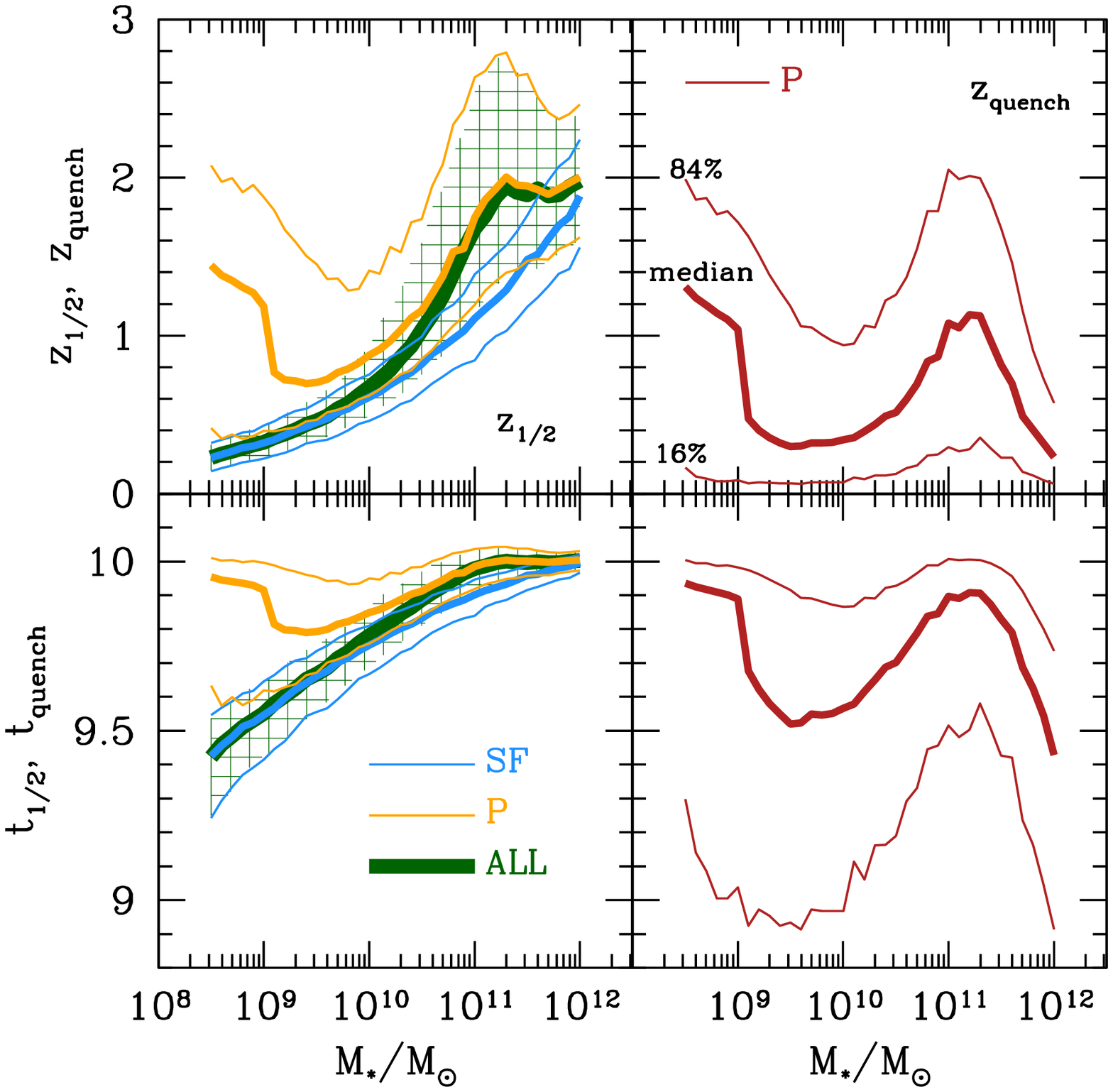}
\caption{\label{fig:zquench} 
{\it{Left}:} The redshift, $\zhalf$, (upper) or lookback time, $t_{1/2}$, (lower) at which a galaxy progenitor reached 50\% of its final stellar mass.  The thick, green line and hatched region shows the median and central 68\% values for a representative population of model galaxies including both star-forming and passive systems, while the thick and thin curves plot the same quantities for only star-forming (blue) or only passive (orange) galaxies at $z=0$.  {\it{Right}:} The redshift (upper) or lookback time (lower) at which passive galaxies at $z=0$ quenched their star formation as a function of $\mnow$.  Thick and thin curves denote median and the central 68\% of values, respectively.  Passive, $10^{8}\,\msun$ galaxies quench early in our model because of the rarity of these objects (see \S\ref{sec:results:ages} of the text).
}
\end{center}
\end{figure*}

\section{Results}\label{sec:results}
We have developed a model to describe galaxy-to-galaxy variations in the buildup and enrichment of stellar mass in terms of the observed star-forming main sequence, the FMR, and the evolving passive galaxy fraction.  In \S\ref{sec:scatt}, we showed how variations in normal star formation, a passive phase, and additional starbursts all lead to scatter in star formation history.  In this section, we explore how these fluctuations contribute to the observed range of ages and stellar metallicities, focusing on measurements at $z\sim0$.

\citet{Gallazzi05} derive average stellar ages and metallicities for $\sim\!175,000$ nearby ($0.05\leq z\leq 0.22$) galaxies from the SDSS Data Release 2 with stellar masses of $\sim 10^{9}$--$10^{12}\,\msun$.  They determine galaxy properties probabilistically by considering a range of star formation histories with bursts superimposed on an assumed exponential decline and using a Bayesian method based on the following observed spectral features: the $4000\,{\rm \AA}$ Balmer break, H$\beta$, H$\delta_{\rm A}$+H$\gamma_{\rm A}$, [Mg$_{2}$ Fe], and [Mg Fe]'.  Because of their reliance on iron features, we correct the metallicities from our oxygen-derived results to account for variations in the $\alpha/{\rm Fe}$ ratio with O/H \citep{Stoll13}, adopting the newest measurements of the solar oxygen abundance \citep{Caffau11}.  

For a given input value of $\mnow$, each run of our method returns a different star formation histories and hence a different galaxy-averaged age and stellar metallicity.  To compare these model results with the \citet{Gallazzi05} observations, we construct samples of $10^{3}$ mock galaxies in mass bins of $\Delta\log\mnow=0.1$ and calculate the median and central 68\% range of the output age and stellar metallicity (see Appendix).  We define the operator $\DD$ to be the range of logarithmic values of the acted-on physical variable such that 16\% of galaxies in the given bin have values above the range and a further 16\% have values below the range.  Thus, $\DD T$ and $\DD \Zstar$, respectively, quantify the range of average ages and stellar metallicities, in dex, of the central 68\% of galaxies.  For example, a value of $\log\DD \Zstar=-1$ implies that the central 68\% of $\Zstar$ spans 0.1~dex.

In addition to the intrinsic scatter, measurement uncertainties also contribute to the observed $\DD T$ and $\DD \Zstar$.  We attempt to compute an approximate correction for this effect by subtracting the uncertainties from the observed scatter in quadrature:
\begin{equation}\label{eq:DXcorr}
\DD X_{\rm{corrected}}=\left[\left(\DD X_{\rm{uncorrected}}\right)^2-\left(2\,\sigX\right)^2\right]^{1/2},
\end{equation}
where $X$ may stand for either $T$ or $\Zstar$.  However, we use equation~\ref{eq:DXcorr} with caution, since none of the relevant fluctuations or uncertainties are Gaussian and the result depends as much on the \citet{Gallazzi05} estimates of their uncertainties as on the data themselves.  Thus, these ``corrected'' values should serve more as a guide to the potential effect of observational uncertainties than a true measure of the intrinsic scatter. 

Finally, we note that, while we implement the quenching of star formation in our MSI formulation using the evolution of the stellar mass function via equation~\ref{eq:Ppswitch}, our results are not explicitly designed to reproduce this evolution.  Thus, a comparison to these observations would represent an additional constraint on our model.  However, such an analysis is beyond the scope of the current study, and we leave it for future work.

%-------------------------------------------------------------------------------------------------------------
\subsection{Galaxy Ages}\label{sec:results:ages}

\begin{figure}
\begin{center}
\includegraphics[width=\columnwidth,trim=15 20 235 0,clip]{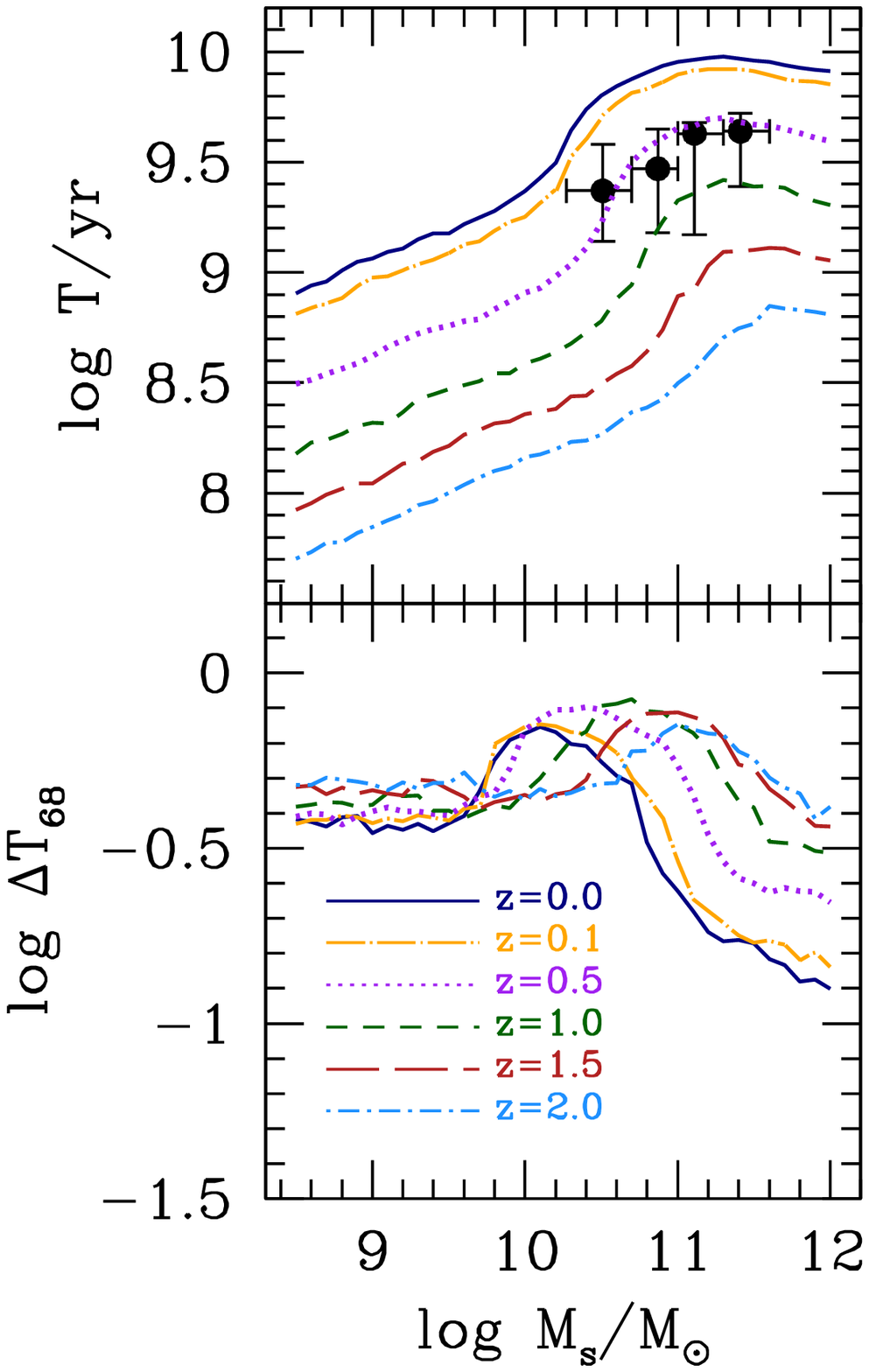}
\caption{\label{fig:T_M_z} 
The median (top panel) and central 68\% interval range (bottom panel) for model luminosity-weighted galactic stellar age as a function of total stellar mass at $z=0$ (solid, dark blue), 0.1 (dot-long-dashed, orange), 0.5 (dotted, purple), 1 (short-dashed, green), 1.5 (long-dashed, red), and 2 (dot-dashed, light blue).  All model results in this plot assume $\Dtc=10^{9}\,\yr$ and include quenching.  For comparison, points and error bars show median age measurements from \citet{Gallazzi14} at $z=0.7$.
}
\end{center}
\end{figure}

We compare our results for galactic ages to observations from \citet[][thick curve and gray shading]{Gallazzi05} in Figure~\ref{fig:scatter1}.  As shown in the left column, the luminosity-weighted ages of star-forming galaxies with $\mnow\gtrsim 10^{10}\,\msun$ are significantly lower than those of the general population.  However, including a passive galaxies in our model, as described in \S\ref{sec:scatt:passive} and shown in the middle and right columns, almost perfectly resolves for the discrepancy.  Moreover, the quenching of star formation produces a characteristic bump in the value of $\DD T$ for the full galaxy population around $10^{10.5}\,\msun$ that is also seen in the observations, where the passive fraction is significant enough to strongly affect a moderate number of galaxies but not so high as to influence nearly all galaxies in the same way.  However, we note that this mechanism results in somewhat less scatter than observed at the highest stellar masses.

Quenching also suppresses the difference between our mass- and luminosity-weighted results.  Among star-forming galaxies we find that luminosity-weighted observations give ages that are significantly younger than the ``true" mass-weighted values, by a factor of $\sim 2$--3, consistent with results from \citet{TS09}.  However, quenching, which affects most galaxies with stellar masses above $\sim 10^{10.5}\,\msun$, removes the more recent stars that would have dominated a luminosity-weighted average and produces agreement between the two weightings.

In Figure~\ref{fig:scatter1}, we also explore the effect of varying the correlation time-scale of normal star formation and show results for $\Dtc=10^{8}$ (short-dashed), $10^{9}$ (solid), and $10^{10}\,\yr$ (long-dashed).  While different values of $\Dtc$ result in nearly identical median ages at a given stellar mass, longer correlation times produce more scatter among star-forming systems from galaxy to galaxy, as expected.  However, in a representative sample of both star-forming and passive systems, quenching appears to be the dominant source of scatter in galaxies with $\mnow\gtrsim 10^{10}\,\msun$ so that the effect of changing $\Dtc$ in this mass range is negligible.  As a consequence, distinguishing between different drivers of star formation based on $\Dtc$ may only be possible with a purely star-forming sample.  Nevertheless, though the observational uncertainties in the age measurements make it difficult to clearly constrain model parameters from the observed values of $\DD T$, our results for dwarf systems seem to prefer $\Dtc>10^{9}\,\yr$.  This may indicate that scatter in star formation histories of star-forming systems owes to more than just small instabilities and clumps in cosmic cold gas accretion operating on short time-scales \citep[e.g.,][]{Genel10, Forbes14}.  Instead, galaxy environment may play a role \citep[e.g.,][]{Peng10, Pasquali10, Lin14}.

Since passive systems are clearly essential to reproducing the full population of galaxies reflected in the \citet{Gallazzi05} observations, we include them in presenting our results in the rest of this work unless specifically noted otherwise.  We also note that, because the median and scatter in galactic ages of massive galaxies is a direct result of the evolving passive fraction, successfully reproducing both sets of data \citep[e.g.,][]{Vogelsberger14} should count only as a single piece of evidence for the success of a theoretical model. 

In addition to the distribution of final galaxy ages, our framework supplies us with full star formation histories from which we can extract interesting characteristic quantities.  The right panel of Figure~\ref{fig:zquench} shows the redshifts, $\zhalf$, (top panels) or lookback times (bottom panels) at which the progenitors of $z=0$ galaxies reached 50\% of their final stellar mass.  Among star-forming galaxies, this redshift increases with increasing stellar mass, but the behavior among galaxies that are quenched by $z=0$ is more complicated.  For example, despite the young average ages of $10^{8.5}\,\msun$ galaxies, systems of this mass that are passive at $z=0$ have a median $\zhalf$ of about $\sim 1.5$, which is a reflection of the very small fraction of $10^{8.5}\,\msun$ galaxies that are quenched.  Because these systems have such low $\fpass$ in our extrapolation of equation~\ref{eq:fpass}, they are more likely to be the descendants of more massive galaxies in which star formation quenches early and recycling reduces the stellar mass down to its final value than the descendants of recently star-forming galaxies with masses close to $10^{8.5}\,\msun$.  This scenario among dwarf galaxies is also suggested by the right panel of Figure~\ref{fig:zquench}, which shows the redshift, $\zq$, at which today's passive galaxies quenched.  The limitations of our model may also be apparent in the declining value of $\zq$ with increasing stellar mass beyond $\sim10^{11}\,\msun$; including mergers in these systems may act to flatten the trend.  However, the behavior of $\zhalf$ for  the total galaxy population is relatively straightforward and reflects our choice of $\fpass$ in equation~\ref{eq:fpass}: median values of $\zhalf$ increase with stellar mass, tracing values for star-forming galaxies for $\mnow\lesssim10^{10}\,\msun$ and those for passive galaxies at $\mnow \gtrsim10^{11}\,\msun$.

Given the excellent agreement between our model results and observations for galactic ages of nearby galaxies, we extend our model to higher redshifts and, in Figure~\ref{fig:T_M_z}, present our calculations of luminosity-weighted ages as a function of stellar mass for $\zf=0$, 0.1, 0.5, 1.0, 1.5, and 2.0.  We demonstrate the validity of these results by demonstrating their excellent agreement with galaxy age measurements at $z=0.7$ by \citet{Gallazzi14}\footnote{Note that the observed sample from \citet{Gallazzi14} is not completely representative of the underlying population, but passive systems general dominate in this mass range.}, but our computations are also in agreements with observations in \citet{Choi14}.  As $\zf$ increases, our results quantify the monotonic decrease in the median galaxy-averaged age and demonstrate an increased scatter in ages at higher redshifts.  Alternatively, as a function of $\mfinal$, our model reveals that the increase in luminosity-weighted age toward higher stellar masses flattens around $\mfinal=10^{11}\,\msun$ at $z=0$ due to quenching and that the flattening begins at progressively higher masses for galaxies at earlier redshifts.  Moreover, the transition from mostly star-forming low-mass systems to mostly-passive massive systems produces a smoother behavior in the $T$--$\mfinal$ at higher redshifts.  While our results at higher redshifts are only as good as our extrapolations of the empirical fitting relations that make up our model, improved mass-complete samples of star formation rates and passive galaxy fractions at $\log\mstar\ll 9$ will allow for additional refinements of our predictions.

%-------------------------------------------------------------------------------------------------------------
\subsection{Stellar Metallicities}\label{sec:results:metals}

\begin{figure*}
\begin{center}
\includegraphics[width=\textwidth,trim=20 40 20 0,clip]{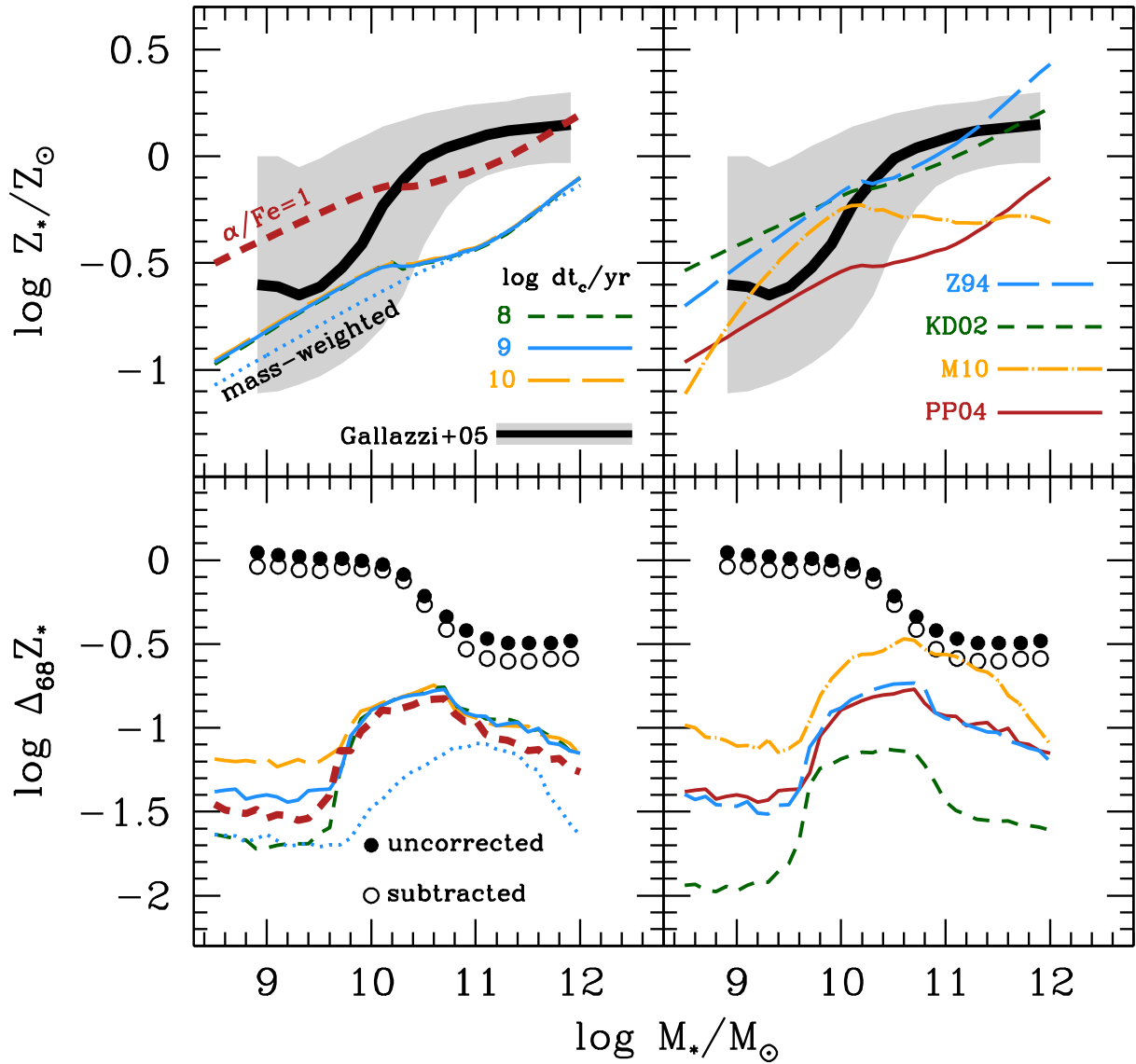}
\caption{\label{fig:scatter3} 
The median (top row) and central 68\% interval range (bottom row) for luminosity-weighted, stellar iron metallicity of a population that includes both passive and star-forming galaxies as a function of stellar mass at $z=0$.  Thick lines and shaded regions in the top row indicate median and scatter in the \citet{Gallazzi05} observations of $\Zstar$ (corrected as discussed in the text), while filled and open circles in the bottom row showing the uncorrected and uncertainty-subtracted central 68\% interval, respectively.  In the left column, short-dashed (green), solid (blue), and long-dashed (orange) curves compare the effect of different correlation time-scales between fluctuations in normal star formation---$\Dtc=10^{8}$, $10^{9}$, and $10^{10}\,\yr$, respectively---assuming the FMR calibration from PP04.  The dotted (blue) curve shows the mass-weighted results, while the thick, short-dashed (red) line demonstrates the effect of ignoring the correction from $\alpha$- to iron-abundance; both assume $\Dtc=10^{9}\,\yr$ and the PP04 calibration.  In the right column, solid (blue), dot-dashed (orange), short-dashed (green), and long-dashed (red) curves compare results from different FMR calibrations of PP04, M10, KD02, and Z94, respectively, with a fiducial correlation time-scale of $\Dtc=10^{9}\,\yr$.
}
\end{center}
\end{figure*}

The favorable comparisons between our results and the observed median and scatter in galactic ages contrast with those for stellar metallicity.  Qualitatively, we generally find higher stellar metallicities in older, more massive galaxies, as observed.   However, we have difficulty reproducing both the median and distribution of stellar metallicities in detail as depicted in Figure~\ref{fig:scatter3}.  The left column shows the effect of varying $\Dtc$ using the FMR indicators of PP04, while the right column assumes $\Dtc=10^{9}\,\yr$ and demonstrates the differences among alternate derivations of the FMR.  

As with the ages in Figure~\ref{fig:scatter1}, all values of $\Dtc$ result in identical median stellar metallicities at all stellar masses.  This is also true for the scatter among galaxies with masses above $\sim 10^{10}\,\msun$, where quenching begins to dominate variations in the star formation history.  

With our fiducial $\alpha$/Fe ratios from \citet{Stoll13}, different FMR calibrations produce results which cluster around the median \citet{Gallazzi05} metallicities for masses less than $\sim10^{10}\,\msun$.  At larger masses, they somewhat underestimate the observed results as a group, but the calibrations of KD02 and Z94 are relatively more successful on their own.  With regard to the metallicity scatter, the Z94 and PP04 calibrations result in values of $\DD \Zstar$ consistent with observations for $\mnow \gtrsim 10^{10.5}\,\msun$; the M10 and KD02 give slightly too much or too little scatter, respectively, in this mass range.  However, all FMR models produce 1--2 orders-of-magnitude less scatter than observed among galaxies with masses below $\sim10^{10.5}\,\msun$.  In principle, we could derive an FMR that produces agreement between our model and observations, but this exercise is complicated by uncertainties in comparing the model and the data (see \S\ref{sec:results:metals:comp}.)

\subsubsection{Uncertainties in the Model--Data Comparison}\label{sec:results:metals:comp}
There are several reasons why we may not have expected to predict the observed metallicity distribution correctly.  We divide these into three categories: unaccounted for physical issues with (1) the metallicity of the gas or (2) translating the oxygen-traced metallicity of gas to the iron-traced metallicity of the stars that form out of that gas and (3) methodological issues with our approach or in the \citet{Gallazzi05} data.  Problems in any of these groups can influence our comparisons of both the median and scatter in metallicity.  

The calibration of the gas-phase metallicity, which we discussed in \S\ref{sec:basic:mh}, is not the only uncertainty affecting our use of the FMR.  Extrapolation is a further cause for concern, particularly at $\mstar \gtrsim 10^{10.5}\,\msun$, beyond the \citet{AM13} sample that was used to fit several of our FMR choices, where \citet{Yates12} find evidence for an turnover in the general relationship between metallicity and the star formation rate.  However, even with a perfect characterization of the $z=0$ FMR across all stellar masses and star formation rates, we would not be able to rule out an evolution in the relation with time.  Although we ignore such evolution, we may actually expect it on theoretical grounds since galactic outflows, which alter the relative amounts of metals and gas a galaxy retains \citep[e.g.,][]{Peeples11}, depend on the evolving galactic potential well \citep[e.g.,][]{Murray05, Murray10}.  Observationally, results are conflicted; some high-redshift studies report metallicities consistent with the local FMR \citep[e.g.,][]{Mannucci10, LaraLopez10, Wuyts12, Henry13a, Henry13b}, while others paint a more uncertain picture \citep[e.g.,][]{Sanchez13, Hughes13, Yabe14, dlReyes14, Sanders14}.  To further complicate these observational tests, the local calibrations of O/H metallicity indicators may not apply in high-redshift galaxies \citep{Kewley13} if the interstellar media of these systems exhibit higher densities and pressures \citep[e.g.][]{MF12, MF13} or harder radiation fields \citep[but see][]{Juneau14}.  Finally, the FMR is not as completely free of scatter \citep{AM13} as we assumed, particularly at high redshift \citep[e.g.,][]{Wuyts12}; like scatter in the $\mstar$--$\sfr$ relation, these fluctuations would imprint on the resulting metallicity distributions from our model.

In addition to the oxygen-traced metallicity in the gas-phase, the iron-traced metallicity of stars formed from that gas is also uncertain, and the largest complicating factor is the $\alpha$/Fe ratio.  As discussed in \S\ref{sec:basic:mh}, we translated the oxygen-weighted gas-phase metallicities of the FMR to stellar metallicities measured relative to iron using the $\alpha$/Fe-O/H relation from \citet{Stoll13} as a fiducial calibration.  Figure~\ref{fig:scatter3} demonstrates that the effect of this conversion is to reduce the median stellar metallicities by a factor of 2--3.  However, the differing delay times between iron-producing Type Ia supernovae (long) and iron- and oxygen-producing Type II supernovae (short) after the formation of a stellar population suggests an underlying age dependence of $\alpha$/Fe \citep{Maoz12}.\footnote{Complicating a pure age dependence, we also note that the different production sites for the two sets of elements could lead to different outflow loading factors and play a role in determining $\alpha$/Fe.}  Thus, the mass-age relation among galaxies (see \S\ref{sec:results:ages}) implies a secondary correlation between $\alpha$/Fe and stellar mass \citep{Thomas05, Arrigoni10}.  We also test (but do not plot) our results using this type of calibration from \citet{Arrigoni10} and find that, while this choice over-produces the metallicity in dwarf galaxies, results from different formulations of the FMR bracket the median observed metallicities above $\sim 10^{10}\,\msun$.  However, the $\mstar$--$\alpha$/Fe relation does not produce the increase in metallicity scatter that we would expect from a true age-$\alpha$/Fe relation because the significant scatter in the mass-age relation is washed out.

There are also several methodological issues that could affect our results and comparisons with observations.  For example, \citet{TS09} demonstrate a potentially important difference between the type of luminosity-weighting that we use and an SSP-weighting, which is more relevant for the \citet{Gallazzi05} spectral index calibrations \citep[see also][]{Panter08}.  Moreover, aperture bias, particularly for galaxies with both bulge and disk components, is partially responsible for the flattening of the mass-stellar metallicity relation in the SDSS measurements \citep{Kauffmann03}, and the choice of model metallicity grids \citep{Koleva08} may also have a subtle effect.  Finally, observational biases may skew the statistics of dwarfs galaxies in the \citet{Gallazzi05} sample away from those of the total population.  In particular, because O and B stars in regions of high star formation rate hinder measurements of the absorption lines used to derive stellar metallicities, the observed sample may have a higher proportion of passive systems than expected from equation~\ref{eq:fpass}.  Although the observed trend in the median metallicity with stellar mass observed by \citet{Gallazzi05} appears to continue in galaxies below the lower SDSS mass limit---around $10^{9}\,\msun$---to Milky Way and M31 satellites with masses as low as $\sim 10^{3}\,\msun$ \citep{Kirby13}, the agreement is not perfect.  Further, the significant amount of scatter in the \citet{Gallazzi05} metallicity measurements at $\sim 10^{9}\,\msun$ is not observed in the \citet{Kirby13} dwarf satellite sample at $\sim10^{8.5}\,\msun$.  This apparent tension in the available data from the literature suggests that the amount of stellar metallicity scatter among dwarf galaxies is not yet a resolved question, though it may simply be an issue of comparing satellites with systems in the field.  New work by \citet{Choi14} analyzing observations specifically of passively evolving galaxies may provide more useful comparisons with our model moving forward \citep[see also][]{Conroy14}.

The reader may also imagine several other factors that could produce the discrepancies between our results and the SDSS metallicity data.  Yet, two possibilities that we do not consider to be significant for our calculation of stellar metallicities are the initial mass function (IMF) of stars and mixing.  While the IMF has a strong impact on the rate at which galaxies produce metals and may vary in certain types of systems \citep{Hopkins12c, CvD12, Conroy13b, Geha13}, our empirical inputs to our model should already reflect these systematic differences.  The only manner by which the IMF can significantly influence our results is through its effect on the observed stellar mass--star formation rate relation, and such a change may alter the good agreement between our results and observed galaxy ages.  On the other hand, mixing may contribute in two ways.  First, unresolved gradients, particularly at higher redshifts \citep[e.g.,][]{Jones13a, Yuan13b}, in the gas-phase metallicity may result in many regions with metallicities significantly different from the measured average.  However, because measurements of the galaxy-averaged $\Zg$ are effectively weighted by star formation rate, the values accurately reflect the gas producing new stars.  Mixing may be a concern either if gas in \hii\ regions is not uniformly mixed or because gas-phase metallicity indicators originate in \hii\ regions rather than the cold, neutral gas where stars.  However, local observations attest to the uniformity of gas-phase metallicities (if not of temperatures) within \hii\ regions \citep[e.g.,][]{ODell03, Garcia-Rojas04} and the agreement between measurements in these regions and the associated stars \citep{SS11, NP12}.  

In summary, it appears that we can resolve the discrepancy between our model metallicity distributions and those observed by \citet{Gallazzi05} through a combination of reconciling the discontinuous scatter between SDSS dwarfs and the satellites of the Milky Way and M31, correctly calibrating the FMR, and implementing an $\alpha$/Fe ratio that depends on the age of the stellar population.

\subsubsection{The Stellar Metallicity--Age Relation}
The last point relating stellar metallicity to age motivates us to consider the relationship between these two output quantities from our model.  In principle, we can calculate $\Zstar$ as a function of $T$ by generating a very large number of star formation histories at a variety of stellar masses, weighted according to the stellar mass function, and compiling the resulting metallicities into bins of galactic age.  Instead, we suggest an analytic treatment to approximate the same result in a much shorter time and with much more physical transparency.  We have additionally carried out the brute-force approach with sufficient accuracy to ensure that this new approach achieves the same result within the uncertainties of our basic model.

To begin, we assume that for each stellar mass at $z=0$, age and stellar metallicity are jointly log-normal.\footnote{While this assumption is not precisely valid, the approximation is sufficient for our purposes.}  The probability distribution of $\log \Zstar$ as a function of $\log T$ is, then,
\begin{equation}\label{eq:ZvsT}
\frac{\dd P(\Zstar,T)}{\dd\log \Zstar} \propto \int\!\! \frac{\dd^2P_{\rm biv}(\Zstar,T|\mstar)}{\dd\log \Zstar\,\dd\log T}\,\frac{\dd n(\mstar, z=0)}{\dd\log \mstar}\,\dd\log \mstar.
\end{equation}
The first factor in the integral is the bivariate, Gaussian probability distribution given by
\begin{equation}\label{eq:Pjoint}
\frac{\dd^2P_{\rm biv}}{\dd x\,\dd y}=\frac{\exp\!\left[\frac{-1}{2\,\left(1-\rho_{xy}^2\right)}\,\left(\frac{x^2}{\sigma_x^2}+\frac{x^2}{\sigma_x^2}-2\,\rho_{xy}\left|\frac{x\,y}{\sigma_x\,\sigma_y}\,\right|\right)\right]}{2\,\pi\,\sigma_{x}\,\sigma_{y}\,\sqrt{1-\rho^2_{xy}}},
\end{equation}
where $x$ and $y$ are jointly normal, mean-subtracted quantities and
\begin{equation}\label{eq:rho}
\rho_{xy}=\frac{\left<x\,y\right>}{\sigma_x\,\sigma_y}
\end{equation}
is their correlation coefficient.  Note that, if $x$ and $y$ are uncorrelated (i.e., $\rho_{xy}=0$), then equation~\ref{eq:Pjoint} reduces to the product of two Gaussians.  We can compute the correlation coefficients between $\log T$ and $\log \Zstar$, as well as the means and variances of these quantities, from the star formation histories that we calculate at the beginning of this section for each stellar mass.  

Because the stellar metallicity distributions from our method do not agree in detail with those observed, we do not present our results for this calculation.  However, we find a general trend of higher metallicities for older galaxies.  We also already see tentative indications that the scatter among metallicities at a given age is less than at a given stellar mass, though the details depend on the specific choice of FMR calibration.  After implementing the solutions to our discrepant metallicities discussed in this section, a more careful analysis of the relationship between $\Zstar$ and $T$ using our framework would be revealing and contribute to an interesting conceptualization of galaxies as inhabiting a stellar mass, age, and stellar metallicity landscape.

%-------------------------------------------------------------------------------------------------------------
%       A Way Forward
%-------------------------------------------------------------------------------------------------------------
\section{Conclusions}\label{sec:forward}

In this study, we develop a cohesive theoretical formalism for translating empirical relations into an understanding of the variations in galactic star formation histories.  We achieve this goal by incorporating into the Main Sequence Integration method the scatter suggested by the evolving fraction of quiescent galaxies and the spread in the observed stellar mass--star formation rate relation.  We find excellent agreement between our model results and the \citet{Gallazzi05} measurements of galactic ages---and approximate agreement with their median stellar metallicities---as a function of stellar mass but are unable to produce the observed variations among the stellar metallicities of dwarf galaxies (see Appendix for information about our tabulated results).

Our work confirms that quenching is the key factor in determining the age distribution of massive systems.  Because our implementation of quenching is stochastic and does not account for the frequency of different environments, environmental effects likely do not contribute very significantly; observational correlations between environment and age may simply result from the strong clustering of massive galaxies into overdense regions.  On the other hand, among star-forming galaxies, variations in star formation rate likely correlate on time-scales longer than $10^{8}\,\yr$---unless additional starbursts strongly influence dwarf galaxy formation \citep[e.g.][but see also \citealt{GarrisonKimmel13}]{Weisz12, Amorisco14, Madau14}---and probably around $10^{9}\,\yr$, which suggests that environmental effects could play some role in setting star formation rates.  However, additional starbursts beyond the normal scatter reflected in observed stellar mass--star formation rate relation are not required to explain the distribution of stellar age as a function of mass.  

Further, our work may reveal missing ingredients in the understanding of galaxy metallicities.  For example, we find that results using representative formulations of the fundamental relation among gas-phase metallicity, stellar mass, and star formation rate approximately bracket the observed median stellar metallicities.  Thus, we conclude that the uncertain calibration of gas-phase metallicity is at least partially responsible for our inability to reproduce the observations in detail.  However, properly modeling the $\alpha$/Fe ratio as a function of the age of a stellar population also promises to improve significantly our comparisons with the data---particularly in generating the stellar metallicity scatter among dwarf galaxies, which retain less than $5\%$ of the metals they have ever made in their stars \citep{Peeples14}---and our understanding of the buildup of the stellar metallicity relation through time.

Moreover, directly connecting empirical phenomenology with our framework for galaxy evolution gives us deeper insights into both data and theory.  For example, we can improve the derivation of galaxy properties from future observations with our MSI-formalism, iteratively solving for the stellar masses and star formation rates, with self-consistent star formation histories.  Such a procedure would yield more accurate results than would be achievable by assuming, e.g., a simple, exponentially-declining star formation rate in the required population synthesis calculation.  Additionally, with regard to theoretical models, our framework points to degeneracies in the testing of numerical simulations and semi-analytic prescriptions against observations and suggests that physical processes correlating star formation on different time-scales may be distinguished by the scatter in the resulting galactic ages of dwarf galaxies.

Thus, our framework clearly provides a useful organizational structure through which the behavior of galaxy properties can answer questions about the star formation and enrichment history of universe.  These answers will increasingly clarify the field of galaxy formation as new streams of data continue to flow.

%-----------------------------------------------------------------------------------------------------------
%       Acknowledgements
%-------------------------------------------------------------------------------------------------------------
\section{Acknowledgements}

We thank Gabe Brammer, Steve Furlanetto, Anna Gallazzi, Evan Kirby, Kai Noeske, Rachel Somerville, David Weinberg, and Kate Whitaker for helpful comments and discussions.  
JAM acknowledges support from the David and Lucile Packard Foundation and NASA grant NNX12AG73G.  MSP acknowledges support from the Southern California Center for Galaxy Evolution, a multi-campus research program funded by the University of California Office of Research.  This research has made extensive use of NASA's Astrophysics Data System.  

%-------------------------------------------------------------------------------------------------------------
%       Bibliography
%-------------------------------------------------------------------------------------------------------------

\bibliography{ms.bbl}

\begin{thebibliography}{93}
\expandafter\ifx\csname natexlab\endcsname\relax\def\natexlab#1{#1}\fi

\bibitem[{{Amorisco} {et~al.}(2014){Amorisco}, {Zavala}, \& {de
  Boer}}]{Amorisco14}
{Amorisco}, N.~C., {Zavala}, J., \& {de Boer}, T.~J.~L. 2014, \apjl, 782, L39

\bibitem[{{Andrews} \& {Martini}(2013)}]{AM13}
{Andrews}, B.~H., \& {Martini}, P. 2013, \apj, 765, 140

\bibitem[{{Arrigoni} {et~al.}(2010){Arrigoni}, {Trager}, {Somerville}, \&
  {Gibson}}]{Arrigoni10}
{Arrigoni}, M., {Trager}, S.~C., {Somerville}, R.~S., \& {Gibson}, B.~K. 2010,
  \mnras, 402, 173

\bibitem[{{Behroozi} {et~al.}(2013){Behroozi}, {Wechsler}, \&
  {Conroy}}]{Behroozi13}
{Behroozi}, P.~S., {Wechsler}, R.~H., \& {Conroy}, C. 2013, \apj, 770, 57

\bibitem[{{Bielby} {et~al.}(2012){Bielby}, {Hudelot}, {McCracken}, {Ilbert},
  {Daddi}, {Le F{\`e}vre}, {Gonzalez-Perez}, {Kneib}, {Marmo}, {Mellier},
  {Salvato}, {Sanders}, \& {Willott}}]{Bielby12}
{Bielby}, R., {Hudelot}, P., {McCracken}, H.~J., {Ilbert}, O., {Daddi}, E., {Le
  F{\`e}vre}, O., {Gonzalez-Perez}, V., {Kneib}, J.-P., {Marmo}, C., {Mellier},
  Y., {Salvato}, M., {Sanders}, D.~B., \& {Willott}, C.~J. 2012, \aap, 545, A23

\bibitem[{{Bigiel} {et~al.}(2008)}]{Bigiel08}
{Bigiel}, F., {et~al.} 2008, \aj, 136, 2846

\bibitem[{{Brammer} {et~al.}(2011){Brammer}, {Whitaker}, {van Dokkum},
  {Marchesini}, {Franx}, {Kriek}, {Labb{\'e}}, {Lee}, {Muzzin}, {Quadri},
  {Rudnick}, \& {Williams}}]{Brammer11}
{Brammer}, G.~B., {Whitaker}, K.~E., {van Dokkum}, P.~G., {Marchesini}, D.,
  {Franx}, M., {Kriek}, M., {Labb{\'e}}, I., {Lee}, K.-S., {Muzzin}, A.,
  {Quadri}, R.~F., {Rudnick}, G., \& {Williams}, R. 2011, \apj, 739, 24

\bibitem[{{Bruzual} \& {Charlot}(2003)}]{BC03}
{Bruzual}, G., \& {Charlot}, S. 2003, \mnras, 344, 1000

\bibitem[{{Caffau} {et~al.}(2011){Caffau}, {Ludwig}, {Steffen}, {Freytag}, \&
  {Bonifacio}}]{Caffau11}
{Caffau}, E., {Ludwig}, H.-G., {Steffen}, M., {Freytag}, B., \& {Bonifacio}, P.
  2011, \solphys, 268, 255

\bibitem[{{Chabrier}(2003)}]{Chabrier03}
{Chabrier}, G. 2003, \pasp, 115, 763

\bibitem[{{Choi} {et~al.}(2014){Choi}, {Conroy}, {Moustakas}, {Graves},
  {Holden}, {Brodwin}, {Brown}, \& {van Dokkum}}]{Choi14}
{Choi}, J., {Conroy}, C., {Moustakas}, J., {Graves}, G.~J., {Holden}, B.~P.,
  {Brodwin}, M., {Brown}, M.~J.~I., \& {van Dokkum}, P.~G. 2014, \apj, 792, 95

\bibitem[{{Conroy} {et~al.}(2013){Conroy}, {Dutton}, {Graves}, {Mendel}, \&
  {van Dokkum}}]{Conroy13b}
{Conroy}, C., {Dutton}, A.~A., {Graves}, G.~J., {Mendel}, J.~T., \& {van
  Dokkum}, P.~G. 2013, \apjl, 776, L26

\bibitem[{{Conroy} {et~al.}(2014){Conroy}, {Graves}, \& {van
  Dokkum}}]{Conroy14}
{Conroy}, C., {Graves}, G.~J., \& {van Dokkum}, P.~G. 2014, \apj, 780, 33

\bibitem[{{Conroy} \& {van Dokkum}(2012)}]{CvD12}
{Conroy}, C., \& {van Dokkum}, P.~G. 2012, \apj, 760, 71

\bibitem[{{Daddi} {et~al.}(2007){Daddi}, {Dickinson}, {Morrison}, {Chary},
  {Cimatti}, {Elbaz}, {Frayer}, {Renzini}, {Pope}, {Alexander}, {Bauer},
  {Giavalisco}, {Huynh}, {Kurk}, \& {Mignoli}}]{Daddi07}
{Daddi}, E., {Dickinson}, M., {Morrison}, G., {Chary}, R., {Cimatti}, A.,
  {Elbaz}, D., {Frayer}, D., {Renzini}, A., {Pope}, A., {Alexander}, D.~M.,
  {Bauer}, F.~E., {Giavalisco}, M., {Huynh}, M., {Kurk}, J., \& {Mignoli}, M.
  2007, \apj, 670, 156

\bibitem[{{de los Reyes} {et~al.}(2014){de los Reyes}, {Ly}, {Lee}, {Salim},
  {Peeples}, {Momcheva}, {Feddersen}, {Dale}, {Ouchi}, {Ono}, \&
  {Finn}}]{dlReyes14}
{de los Reyes}, M.~A., {Ly}, C., {Lee}, J.~C., {Salim}, S., {Peeples}, M.~S.,
  {Momcheva}, I., {Feddersen}, J., {Dale}, D.~A., {Ouchi}, M., {Ono}, Y., \&
  {Finn}, R. 2014, astro-ph/1410.1551

\bibitem[{{Dopita} {et~al.}(2013){Dopita}, {Sutherland}, {Nicholls}, {Kewley},
  \& {Vogt}}]{Dopita13}
{Dopita}, M.~A., {Sutherland}, R.~S., {Nicholls}, D.~C., {Kewley}, L.~J., \&
  {Vogt}, F.~P.~A. 2013, \apjs, 208, 10

\bibitem[{{Forbes} {et~al.}(2014){Forbes}, {Krumholz}, {Burkert}, \&
  {Dekel}}]{Forbes14}
{Forbes}, J.~C., {Krumholz}, M.~R., {Burkert}, A., \& {Dekel}, A. 2014, \mnras,
  443, 168

\bibitem[{{Gallazzi} {et~al.}(2014){Gallazzi}, {Bell}, {Zibetti}, {Brinchmann},
  \& {Kelson}}]{Gallazzi14}
{Gallazzi}, A., {Bell}, E.~F., {Zibetti}, S., {Brinchmann}, J., \& {Kelson},
  D.~D. 2014, \apj, 788, 72

\bibitem[{{Gallazzi} {et~al.}(2005){Gallazzi}, {Charlot}, {Brinchmann},
  {White}, \& {Tremonti}}]{Gallazzi05}
{Gallazzi}, A., {Charlot}, S., {Brinchmann}, J., {White}, S.~D.~M., \&
  {Tremonti}, C.~A. 2005, \mnras, 362, 41

\bibitem[{{Garc{\'{\i}}a-Rojas} {et~al.}(2004){Garc{\'{\i}}a-Rojas}, {Esteban},
  {Peimbert}, {Rodr{\'{\i}}guez}, {Ruiz}, \& {Peimbert}}]{Garcia-Rojas04}
{Garc{\'{\i}}a-Rojas}, J., {Esteban}, C., {Peimbert}, M., {Rodr{\'{\i}}guez},
  M., {Ruiz}, M.~T., \& {Peimbert}, A. 2004, \apjs, 153, 501

\bibitem[{{Garrison-Kimmel} {et~al.}(2013){Garrison-Kimmel}, {Rocha},
  {Boylan-Kolchin}, {Bullock}, \& {Lally}}]{GarrisonKimmel13}
{Garrison-Kimmel}, S., {Rocha}, M., {Boylan-Kolchin}, M., {Bullock}, J.~S., \&
  {Lally}, J. 2013, \mnras, 433, 3539

\bibitem[{{Geha} {et~al.}(2013){Geha}, {Brown}, {Tumlinson}, {Kalirai},
  {Simon}, {Kirby}, {VandenBerg}, {Mu{\~n}oz}, {Avila}, {Guhathakurta}, \&
  {Ferguson}}]{Geha13}
{Geha}, M., {Brown}, T.~M., {Tumlinson}, J., {Kalirai}, J.~S., {Simon}, J.~D.,
  {Kirby}, E.~N., {VandenBerg}, D.~A., {Mu{\~n}oz}, R.~R., {Avila}, R.~J.,
  {Guhathakurta}, P., \& {Ferguson}, H.~C. 2013, \apj, 771, 29

\bibitem[{{Genel} {et~al.}(2010){Genel}, {Bouch{\'e}}, {Naab}, {Sternberg}, \&
  {Genzel}}]{Genel10}
{Genel}, S., {Bouch{\'e}}, N., {Naab}, T., {Sternberg}, A., \& {Genzel}, R.
  2010, \apj, 719, 229

\bibitem[{{Gonz{\'a}lez} {et~al.}(2014){Gonz{\'a}lez}, {Bouwens},
  {Illingworth}, {Labb{\'e}}, {Oesch}, {Franx}, \& {Magee}}]{Gonzalez14}
{Gonz{\'a}lez}, V., {Bouwens}, R., {Illingworth}, G., {Labb{\'e}}, I., {Oesch},
  P., {Franx}, M., \& {Magee}, D. 2014, \apj, 781, 34

\bibitem[{{Gonz{\'a}lez} {et~al.}(2011){Gonz{\'a}lez}, {Labb{\'e}}, {Bouwens},
  {Illingworth}, {Franx}, \& {Kriek}}]{Gonzalez11}
{Gonz{\'a}lez}, V., {Labb{\'e}}, I., {Bouwens}, R.~J., {Illingworth}, G.,
  {Franx}, M., \& {Kriek}, M. 2011, \apjl, 735, L34

\bibitem[{{Guo} {et~al.}(2013){Guo}, {Zheng}, \& {Fu}}]{Guo13}
{Guo}, K., {Zheng}, X.~Z., \& {Fu}, H. 2013, \apj, 778, 23

\bibitem[{{Henry} {et~al.}(2013{\natexlab{a}}){Henry}, {Martin}, {Finlator}, \&
  {Dressler}}]{Henry13a}
{Henry}, A., {Martin}, C.~L., {Finlator}, K., \& {Dressler}, A.
  2013{\natexlab{a}}, \apj, 769, 148

\bibitem[{{Henry} {et~al.}(2013{\natexlab{b}}){Henry}, {Scarlata},
  {Dom{\'{\i}}nguez}, {Malkan}, {Martin}, {Siana}, {Atek}, {Bedregal},
  {Colbert}, {Rafelski}, {Ross}, {Teplitz}, {Bunker}, {Dressler}, {Hathi},
  {Masters}, {McCarthy}, \& {Straughn}}]{Henry13b}
{Henry}, A., {Scarlata}, C., {Dom{\'{\i}}nguez}, A., {Malkan}, M., {Martin},
  C.~L., {Siana}, B., {Atek}, H., {Bedregal}, A.~G., {Colbert}, J.~W.,
  {Rafelski}, M., {Ross}, N., {Teplitz}, H., {Bunker}, A.~J., {Dressler}, A.,
  {Hathi}, N., {Masters}, D., {McCarthy}, P., \& {Straughn}, A.
  2013{\natexlab{b}}, \apjl, 776, L27

\bibitem[{{Hopkins}(2012)}]{Hopkins12c}
{Hopkins}, P.~F. 2012, \mnras, 423, 2037

\bibitem[{{Hughes} {et~al.}(2013){Hughes}, {Cortese}, {Boselli}, {Gavazzi}, \&
  {Davies}}]{Hughes13}
{Hughes}, T.~M., {Cortese}, L., {Boselli}, A., {Gavazzi}, G., \& {Davies},
  J.~I. 2013, \aap, 550, A115

\bibitem[{{Jones} {et~al.}(2013){Jones}, {Ellis}, {Richard}, \&
  {Jullo}}]{Jones13a}
{Jones}, T., {Ellis}, R.~S., {Richard}, J., \& {Jullo}, E. 2013, \apj, 765, 48

\bibitem[{{Jones} {et~al.}(2014){Jones}, {Wang}, {Schmidt}, {Treu}, {Brammer},
  {Bradac}, {Dressler}, {Henry}, {Malkan}, {Pentericci}, \& {Trenti}}]{Jones14}
{Jones}, T., {Wang}, X., {Schmidt}, K., {Treu}, T., {Brammer}, G., {Bradac},
  M., {Dressler}, A., {Henry}, A., {Malkan}, M., {Pentericci}, L., \& {Trenti},
  M. 2014, astro-ph/1410.0967

\bibitem[{{Juneau} {et~al.}(2014){Juneau}, {Bournaud}, {Charlot}, {Daddi},
  {Elbaz}, {Trump}, {Brinchmann}, {Dickinson}, {Duc}, {Gobat}, {Jean-Baptiste},
  {Le Floc'h}, {Lehnert}, {Pacifici}, {Pannella}, \& {Schreiber}}]{Juneau14}
{Juneau}, S., {Bournaud}, F., {Charlot}, S., {Daddi}, E., {Elbaz}, D., {Trump},
  J.~R., {Brinchmann}, J., {Dickinson}, M., {Duc}, P.-A., {Gobat}, R.,
  {Jean-Baptiste}, I., {Le Floc'h}, {\'E}., {Lehnert}, M.~D., {Pacifici}, C.,
  {Pannella}, M., \& {Schreiber}, C. 2014, \apj, 788, 88

\bibitem[{{Karim} {et~al.}(2011){Karim}, {Schinnerer},
  {Mart{\'{\i}}nez-Sansigre}, {Sargent}, {van der Wel}, {Rix}, {Ilbert},
  {Smol{\v c}i{\'c}}, {Carilli}, {Pannella}, {Koekemoer}, {Bell}, \&
  {Salvato}}]{Karim11}
{Karim}, A., {Schinnerer}, E., {Mart{\'{\i}}nez-Sansigre}, A., {Sargent},
  M.~T., {van der Wel}, A., {Rix}, H.-W., {Ilbert}, O., {Smol{\v c}i{\'c}}, V.,
  {Carilli}, C., {Pannella}, M., {Koekemoer}, A.~M., {Bell}, E.~F., \&
  {Salvato}, M. 2011, \apj, 730, 61

\bibitem[{{Kauffmann} {et~al.}(2003){Kauffmann}, {Heckman}, {White}, {Charlot},
  {Tremonti}, {Peng}, {Seibert}, {Brinkmann}, {Nichol}, {SubbaRao}, \&
  {York}}]{Kauffmann03}
{Kauffmann}, G., {Heckman}, T.~M., {White}, S.~D.~M., {Charlot}, S.,
  {Tremonti}, C., {Peng}, E.~W., {Seibert}, M., {Brinkmann}, J., {Nichol},
  R.~C., {SubbaRao}, M., \& {York}, D. 2003, \mnras, 341, 54

\bibitem[{{Kewley} \& {Dopita}(2002)}]{KD02}
{Kewley}, L.~J., \& {Dopita}, M.~A. 2002, \apjs, 142, 35

\bibitem[{{Kewley} {et~al.}(2013){Kewley}, {Dopita}, {Leitherer}, {Dav{\'e}},
  {Yuan}, {Allen}, {Groves}, \& {Sutherland}}]{Kewley13}
{Kewley}, L.~J., {Dopita}, M.~A., {Leitherer}, C., {Dav{\'e}}, R., {Yuan}, T.,
  {Allen}, M., {Groves}, B., \& {Sutherland}, R. 2013, \apj, 774, 100

\bibitem[{{Kewley} \& {Ellison}(2008)}]{KE08}
{Kewley}, L.~J., \& {Ellison}, S.~L. 2008, \apj, 681, 1183

\bibitem[{{Kirby} {et~al.}(2013){Kirby}, {Cohen}, {Guhathakurta}, {Cheng},
  {Bullock}, \& {Gallazzi}}]{Kirby13}
{Kirby}, E.~N., {Cohen}, J.~G., {Guhathakurta}, P., {Cheng}, L., {Bullock},
  J.~S., \& {Gallazzi}, A. 2013, \apj, 779, 102

\bibitem[{{Klypin} {et~al.}(2002){Klypin}, {Zhao}, \& {Somerville}}]{Klypin02}
{Klypin}, A., {Zhao}, H., \& {Somerville}, R.~S. 2002, \apj, 573, 597

\bibitem[{{Koleva} {et~al.}(2008){Koleva}, {Prugniel}, {Ocvirk}, {Le Borgne},
  \& {Soubiran}}]{Koleva08}
{Koleva}, M., {Prugniel}, P., {Ocvirk}, P., {Le Borgne}, D., \& {Soubiran}, C.
  2008, \mnras, 385, 1998

\bibitem[{{Kravtsov} {et~al.}(2004){Kravtsov}, {Gnedin}, \&
  {Klypin}}]{Kravtsov04}
{Kravtsov}, A.~V., {Gnedin}, O.~Y., \& {Klypin}, A.~A. 2004, \apj, 609, 482

\bibitem[{{Lara-L{\'o}pez} {et~al.}(2010){Lara-L{\'o}pez}, {Cepa},
  {Bongiovanni}, {P{\'e}rez Garc{\'{\i}}a}, {Ederoclite}, {Casta{\~n}eda},
  {Fern{\'a}ndez Lorenzo}, {Povi{\'c}}, \& {S{\'a}nchez-Portal}}]{LaraLopez10}
{Lara-L{\'o}pez}, M.~A., {Cepa}, J., {Bongiovanni}, A., {P{\'e}rez
  Garc{\'{\i}}a}, A.~M., {Ederoclite}, A., {Casta{\~n}eda}, H., {Fern{\'a}ndez
  Lorenzo}, M., {Povi{\'c}}, M., \& {S{\'a}nchez-Portal}, M. 2010, \aap, 521,
  L53

\bibitem[{{Leitherer} {et~al.}(1999){Leitherer}, {Schaerer}, {Goldader},
  {Gonz{\'a}lez Delgado}, {Robert}, {Kune}, {de Mello}, {Devost}, \&
  {Heckman}}]{Leitherer99}
{Leitherer}, C., {Schaerer}, D., {Goldader}, J.~D., {Gonz{\'a}lez Delgado},
  R.~M., {Robert}, C., {Kune}, D.~F., {de Mello}, D.~F., {Devost}, D., \&
  {Heckman}, T.~M. 1999, \apjs, 123, 3

\bibitem[{{Leitner}(2012)}]{Leitner12}
{Leitner}, S.~N. 2012, \apj, 745, 149

\bibitem[{{Leitner} \& {Kravtsov}(2011)}]{LK11}
{Leitner}, S.~N., \& {Kravtsov}, A.~V. 2011, \apj, 734, 48

\bibitem[{{Lin} {et~al.}(2014){Lin}, {Jian}, {Foucaud}, {Norberg}, {Bower},
  {Cole}, {Arnalte-Mur}, {Chen}, {Coupon}, {Hsieh}, {Heinis}, {Phleps}, {Chen},
  {Lee}, {Burgett}, {Chambers}, {Denneau}, {Draper}, {Flewelling}, {Hodapp},
  {Huber}, {Kaiser}, {Kudritzki}, {Magnier}, {Metcalfe}, {Price}, {Tonry},
  {Wainscoat}, \& {Waters}}]{Lin14}
{Lin}, L., {Jian}, H.-Y., {Foucaud}, S., {Norberg}, P., {Bower}, R.~G., {Cole},
  S., {Arnalte-Mur}, P., {Chen}, C.-W., {Coupon}, J., {Hsieh}, B.-C., {Heinis},
  S., {Phleps}, S., {Chen}, W.-P., {Lee}, C.-H., {Burgett}, W., {Chambers},
  K.~C., {Denneau}, L., {Draper}, P., {Flewelling}, H., {Hodapp}, K.~W.,
  {Huber}, M.~E., {Kaiser}, N., {Kudritzki}, R.-P., {Magnier}, E.~A.,
  {Metcalfe}, N., {Price}, P.~A., {Tonry}, J.~L., {Wainscoat}, R.~J., \&
  {Waters}, C. 2014, \apj, 782, 33

\bibitem[{{Madau} {et~al.}(2014){Madau}, {Shen}, \& {Governato}}]{Madau14}
{Madau}, P., {Shen}, S., \& {Governato}, F. 2014, \apjl, 789, L17

\bibitem[{{Mannucci} {et~al.}(2010){Mannucci}, {Cresci}, {Maiolino}, {Marconi},
  \& {Gnerucci}}]{Mannucci10}
{Mannucci}, F., {Cresci}, G., {Maiolino}, R., {Marconi}, A., \& {Gnerucci}, A.
  2010, \mnras, 408, 2115

\bibitem[{{Maoz} {et~al.}(2012){Maoz}, {Mannucci}, \& {Brandt}}]{Maoz12}
{Maoz}, D., {Mannucci}, F., \& {Brandt}, T.~D. 2012, \mnras, 426, 3282

\bibitem[{{Mu{\~n}oz}(2012)}]{Munoz12}
{Mu{\~n}oz}, J.~A. 2012, \jcap, 4, 15

\bibitem[{{Mu{\~n}oz} \& {Furlanetto}(2012)}]{MF12}
{Mu{\~n}oz}, J.~A., \& {Furlanetto}, S.~R. 2012, \mnras, 426, 3477

\bibitem[{{Mu{\~n}oz} \& {Furlanetto}(2013)}]{MF13}
---. 2013, \mnras, 435, 2676

\bibitem[{{Mu{\~n}oz} \& {Loeb}(2011)}]{ML11}
{Mu{\~n}oz}, J.~A., \& {Loeb}, A. 2011, \apj, 729, 99

\bibitem[{{Mu{\~n}oz} {et~al.}(2009){Mu{\~n}oz}, {Madau}, {Loeb}, \&
  {Diemand}}]{Munoz09}
{Mu{\~n}oz}, J.~A., {Madau}, P., {Loeb}, A., \& {Diemand}, J. 2009, \mnras,
  400, 1593

\bibitem[{{Murray} {et~al.}(2005){Murray}, {Quataert}, \&
  {Thompson}}]{Murray05}
{Murray}, N., {Quataert}, E., \& {Thompson}, T.~A. 2005, \apj, 618, 569

\bibitem[{{Murray} {et~al.}(2010){Murray}, {Quataert}, \&
  {Thompson}}]{Murray10}
---. 2010, \apj, 709, 191

\bibitem[{{Muzzin} {et~al.}(2013){Muzzin}, {Marchesini}, {Stefanon}, {Franx},
  {McCracken}, {Milvang-Jensen}, {Dunlop}, {Fynbo}, {Brammer}, {Labb{\'e}}, \&
  {van Dokkum}}]{Muzzin13}
{Muzzin}, A., {Marchesini}, D., {Stefanon}, M., {Franx}, M., {McCracken},
  H.~J., {Milvang-Jensen}, B., {Dunlop}, J.~S., {Fynbo}, J.~P.~U., {Brammer},
  G., {Labb{\'e}}, I., \& {van Dokkum}, P.~G. 2013, \apj, 777, 18

\bibitem[{{Nicholls} {et~al.}(2012){Nicholls}, {Dopita}, \&
  {Sutherland}}]{Nicholls12}
{Nicholls}, D.~C., {Dopita}, M.~A., \& {Sutherland}, R.~S. 2012, \apj, 752, 148

\bibitem[{{Nieva} \& {Przybilla}(2012)}]{NP12}
{Nieva}, M.-F., \& {Przybilla}, N. 2012, \aap, 539, A143

\bibitem[{{Noeske} {et~al.}(2007){Noeske}, {Weiner}, {Faber}, {Papovich},
  {Koo}, {Somerville}, {Bundy}, {Conselice}, {Newman}, {Schiminovich}, {Le
  Floc'h}, {Coil}, {Rieke}, {Lotz}, {Primack}, {Barmby}, {Cooper}, {Davis},
  {Ellis}, {Fazio}, {Guhathakurta}, {Huang}, {Kassin}, {Martin}, {Phillips},
  {Rich}, {Small}, {Willmer}, \& {Wilson}}]{Noeske07}
{Noeske}, K.~G., {Weiner}, B.~J., {Faber}, S.~M., {Papovich}, C., {Koo}, D.~C.,
  {Somerville}, R.~S., {Bundy}, K., {Conselice}, C.~J., {Newman}, J.~A.,
  {Schiminovich}, D., {Le Floc'h}, E., {Coil}, A.~L., {Rieke}, G.~H., {Lotz},
  J.~M., {Primack}, J.~R., {Barmby}, P., {Cooper}, M.~C., {Davis}, M., {Ellis},
  R.~S., {Fazio}, G.~G., {Guhathakurta}, P., {Huang}, J., {Kassin}, S.~A.,
  {Martin}, D.~C., {Phillips}, A.~C., {Rich}, R.~M., {Small}, T.~A., {Willmer},
  C.~N.~A., \& {Wilson}, G. 2007, \apjl, 660, L43

\bibitem[{{O'Dell} {et~al.}(2003){O'Dell}, {Peimbert}, \& {Peimbert}}]{ODell03}
{O'Dell}, C.~R., {Peimbert}, M., \& {Peimbert}, A. 2003, \aj, 125, 2590

\bibitem[{{Panter} {et~al.}(2008){Panter}, {Jimenez}, {Heavens}, \&
  {Charlot}}]{Panter08}
{Panter}, B., {Jimenez}, R., {Heavens}, A.~F., \& {Charlot}, S. 2008, \mnras,
  391, 1117

\bibitem[{{Pasquali} {et~al.}(2010){Pasquali}, {Gallazzi}, {Fontanot}, {van den
  Bosch}, {De Lucia}, {Mo}, \& {Yang}}]{Pasquali10}
{Pasquali}, A., {Gallazzi}, A., {Fontanot}, F., {van den Bosch}, F.~C., {De
  Lucia}, G., {Mo}, H.~J., \& {Yang}, X. 2010, \mnras, 407, 937

\bibitem[{{Pe{\~n}a-Guerrero} {et~al.}(2012){Pe{\~n}a-Guerrero}, {Peimbert}, \&
  {Peimbert}}]{Pena12}
{Pe{\~n}a-Guerrero}, M.~A., {Peimbert}, A., \& {Peimbert}, M. 2012, \apjl, 756,
  L14

\bibitem[{{Peeples} \& {Shankar}(2011)}]{Peeples11}
{Peeples}, M.~S., \& {Shankar}, F. 2011, \mnras, 417, 2962

\bibitem[{{Peeples} \& {Somerville}(2013)}]{PS13}
{Peeples}, M.~S., \& {Somerville}, R.~S. 2013, \mnras, 428, 1766

\bibitem[{{Peeples} {et~al.}(2014){Peeples}, {Werk}, {Tumlinson},
  {Oppenheimer}, {Prochaska}, {Katz}, \& {Weinberg}}]{Peeples14}
{Peeples}, M.~S., {Werk}, J.~K., {Tumlinson}, J., {Oppenheimer}, B.~D.,
  {Prochaska}, J.~X., {Katz}, N., \& {Weinberg}, D.~H. 2014, \apj, 786, 54

\bibitem[{{Peng} {et~al.}(2010){Peng}, {Lilly}, {Kova{\v c}}, {Bolzonella},
  {Pozzetti}, {Renzini}, {Zamorani}, {Ilbert}, {Knobel}, {Iovino}, {Maier},
  {Cucciati}, {Tasca}, {Carollo}, {Silverman}, {Kampczyk}, {de Ravel},
  {Sanders}, {Scoville}, {Contini}, {Mainieri}, {Scodeggio}, {Kneib}, {Le
  F{\`e}vre}, {Bardelli}, {Bongiorno}, {Caputi}, {Coppa}, {de la Torre},
  {Franzetti}, {Garilli}, {Lamareille}, {Le Borgne}, {Le Brun}, {Mignoli},
  {Perez Montero}, {Pello}, {Ricciardelli}, {Tanaka}, {Tresse}, {Vergani},
  {Welikala}, {Zucca}, {Oesch}, {Abbas}, {Barnes}, {Bordoloi}, {Bottini},
  {Cappi}, {Cassata}, {Cimatti}, {Fumana}, {Hasinger}, {Koekemoer},
  {Leauthaud}, {Maccagni}, {Marinoni}, {McCracken}, {Memeo}, {Meneux}, {Nair},
  {Porciani}, {Presotto}, \& {Scaramella}}]{Peng10}
{Peng}, Y.-j., {Lilly}, S.~J., {Kova{\v c}}, K., {Bolzonella}, M., {Pozzetti},
  L., {Renzini}, A., {Zamorani}, G., {Ilbert}, O., {Knobel}, C., {Iovino}, A.,
  {Maier}, C., {Cucciati}, O., {Tasca}, L., {Carollo}, C.~M., {Silverman}, J.,
  {Kampczyk}, P., {de Ravel}, L., {Sanders}, D., {Scoville}, N., {Contini}, T.,
  {Mainieri}, V., {Scodeggio}, M., {Kneib}, J.-P., {Le F{\`e}vre}, O.,
  {Bardelli}, S., {Bongiorno}, A., {Caputi}, K., {Coppa}, G., {de la Torre},
  S., {Franzetti}, P., {Garilli}, B., {Lamareille}, F., {Le Borgne}, J.-F., {Le
  Brun}, V., {Mignoli}, M., {Perez Montero}, E., {Pello}, R., {Ricciardelli},
  E., {Tanaka}, M., {Tresse}, L., {Vergani}, D., {Welikala}, N., {Zucca}, E.,
  {Oesch}, P., {Abbas}, U., {Barnes}, L., {Bordoloi}, R., {Bottini}, D.,
  {Cappi}, A., {Cassata}, P., {Cimatti}, A., {Fumana}, M., {Hasinger}, G.,
  {Koekemoer}, A., {Leauthaud}, A., {Maccagni}, D., {Marinoni}, C.,
  {McCracken}, H., {Memeo}, P., {Meneux}, B., {Nair}, P., {Porciani}, C.,
  {Presotto}, V., \& {Scaramella}, R. 2010, \apj, 721, 193

\bibitem[{{Pettini} \& {Pagel}(2004)}]{PP04}
{Pettini}, M., \& {Pagel}, B.~E.~J. 2004, \mnras, 348, L59

\bibitem[{{S{\'a}nchez} {et~al.}(2013){S{\'a}nchez}, {Rosales-Ortega},
  {Jungwiert}, {Iglesias-P{\'a}ramo}, {V{\'{\i}}lchez}, {Marino}, {Walcher},
  {Husemann}, {Mast}, {Monreal-Ibero}, {Cid Fernandes}, {P{\'e}rez},
  {Gonz{\'a}lez Delgado}, {Garc{\'{\i}}a-Benito}, {Galbany}, {van de Ven},
  {Jahnke}, {Flores}, {Bland-Hawthorn}, {L{\'o}pez-S{\'a}nchez}, {Stanishev},
  {Miralles-Caballero}, {D{\'{\i}}az}, {S{\'a}nchez-Blazquez}, {Moll{\'a}},
  {Gallazzi}, {Papaderos}, {Gomes}, {Gruel}, {P{\'e}rez}, {Ruiz-Lara},
  {Florido}, {de Lorenzo-C{\'a}ceres}, {Mendez-Abreu}, {Kehrig}, {Roth},
  {Ziegler}, {Alves}, {Wisotzki}, {Kupko}, {Quirrenbach}, {Bomans}, \& {Califa
  Collaboration}}]{Sanchez13}
{S{\'a}nchez}, S.~F., {Rosales-Ortega}, F.~F., {Jungwiert}, B.,
  {Iglesias-P{\'a}ramo}, J., {V{\'{\i}}lchez}, J.~M., {Marino}, R.~A.,
  {Walcher}, C.~J., {Husemann}, B., {Mast}, D., {Monreal-Ibero}, A., {Cid
  Fernandes}, R., {P{\'e}rez}, E., {Gonz{\'a}lez Delgado}, R.,
  {Garc{\'{\i}}a-Benito}, R., {Galbany}, L., {van de Ven}, G., {Jahnke}, K.,
  {Flores}, H., {Bland-Hawthorn}, J., {L{\'o}pez-S{\'a}nchez}, A.~R.,
  {Stanishev}, V., {Miralles-Caballero}, D., {D{\'{\i}}az}, A.~I.,
  {S{\'a}nchez-Blazquez}, P., {Moll{\'a}}, M., {Gallazzi}, A., {Papaderos}, P.,
  {Gomes}, J.~M., {Gruel}, N., {P{\'e}rez}, I., {Ruiz-Lara}, T., {Florido}, E.,
  {de Lorenzo-C{\'a}ceres}, A., {Mendez-Abreu}, J., {Kehrig}, C., {Roth},
  M.~M., {Ziegler}, B., {Alves}, J., {Wisotzki}, L., {Kupko}, D.,
  {Quirrenbach}, A., {Bomans}, D., \& {Califa Collaboration}. 2013, \aap, 554,
  A58

\bibitem[{{Sanders} {et~al.}(2014){Sanders}, {Shapley}, {Kriek}, {Reddy},
  {Freeman}, {Coil}, {Siana}, {Mobasher}, {Shivaei}, {Price}, \& {de
  Groot}}]{Sanders14}
{Sanders}, R.~L., {Shapley}, A.~E., {Kriek}, M., {Reddy}, N.~A., {Freeman},
  W.~R., {Coil}, A.~L., {Siana}, B., {Mobasher}, B., {Shivaei}, I., {Price},
  S.~H., \& {de Groot}, L. 2014, astro-ph:1408.2521

\bibitem[{{Schaerer} {et~al.}(2013){Schaerer}, {de Barros}, \&
  {Sklias}}]{Schaerer13}
{Schaerer}, D., {de Barros}, S., \& {Sklias}, P. 2013, \aap, 549, A4

\bibitem[{{Schechter}(1976)}]{Schechter76}
{Schechter}, P. 1976, \apj, 203, 297

\bibitem[{{Sim{\'o}n-D{\'{\i}}az} \& {Stasi{\'n}ska}(2011)}]{SS11}
{Sim{\'o}n-D{\'{\i}}az}, S., \& {Stasi{\'n}ska}, G. 2011, \aap, 526, A48

\bibitem[{{Stoll} {et~al.}(2013){Stoll}, {Prieto}, {Stanek}, \&
  {Pogge}}]{Stoll13}
{Stoll}, R., {Prieto}, J.~L., {Stanek}, K.~Z., \& {Pogge}, R.~W. 2013, \apj,
  773, 12

\bibitem[{{Thomas} {et~al.}(2005){Thomas}, {Maraston}, {Bender}, \& {Mendes de
  Oliveira}}]{Thomas05}
{Thomas}, D., {Maraston}, C., {Bender}, R., \& {Mendes de Oliveira}, C. 2005,
  \apj, 621, 673

\bibitem[{{Tinsley}(1975)}]{Tinsley75}
{Tinsley}, B.~M. 1975, \apj, 197, 159

\bibitem[{{Tinsley}(1980)}]{Tinsley80}
---. 1980, \fcp, 5, 287

\bibitem[{{Trager} \& {Somerville}(2009)}]{TS09}
{Trager}, S.~C., \& {Somerville}, R.~S. 2009, \mnras, 395, 608

\bibitem[{{Vogelsberger} {et~al.}(2014){Vogelsberger}, {Genel}, {Springel},
  {Torrey}, {Sijacki}, {Xu}, {Snyder}, {Nelson}, \&
  {Hernquist}}]{Vogelsberger14}
{Vogelsberger}, M., {Genel}, S., {Springel}, V., {Torrey}, P., {Sijacki}, D.,
  {Xu}, D., {Snyder}, G., {Nelson}, D., \& {Hernquist}, L. 2014, \mnras, 444,
  1518

\bibitem[{{Weisz} {et~al.}(2012){Weisz}, {Johnson}, {Johnson}, {Skillman},
  {Lee}, {Kennicutt}, {Calzetti}, {van Zee}, {Bothwell}, {Dalcanton}, {Dale},
  \& {Williams}}]{Weisz12}
{Weisz}, D.~R., {Johnson}, B.~D., {Johnson}, L.~C., {Skillman}, E.~D., {Lee},
  J.~C., {Kennicutt}, R.~C., {Calzetti}, D., {van Zee}, L., {Bothwell}, M.~S.,
  {Dalcanton}, J.~J., {Dale}, D.~A., \& {Williams}, B.~F. 2012, \apj, 744, 44

\bibitem[{{Whitaker} {et~al.}(2011){Whitaker}, {Labb{\'e}}, {van Dokkum},
  {Brammer}, {Kriek}, {Marchesini}, {Quadri}, {Franx}, {Muzzin}, {Williams},
  {Bezanson}, {Illingworth}, {Lee}, {Lundgren}, {Nelson}, {Rudnick}, {Tal}, \&
  {Wake}}]{Whitaker11}
{Whitaker}, K.~E., {Labb{\'e}}, I., {van Dokkum}, P.~G., {Brammer}, G.,
  {Kriek}, M., {Marchesini}, D., {Quadri}, R.~F., {Franx}, M., {Muzzin}, A.,
  {Williams}, R.~J., {Bezanson}, R., {Illingworth}, G.~D., {Lee}, K.-S.,
  {Lundgren}, B., {Nelson}, E.~J., {Rudnick}, G., {Tal}, T., \& {Wake}, D.~A.
  2011, \apj, 735, 86

\bibitem[{{Whitaker} {et~al.}(2012){Whitaker}, {van Dokkum}, {Brammer}, \&
  {Franx}}]{Whitaker12}
{Whitaker}, K.~E., {van Dokkum}, P.~G., {Brammer}, G., \& {Franx}, M. 2012,
  \apjl, 754, L29

\bibitem[{{Wuyts} {et~al.}(2012){Wuyts}, {Rigby}, {Sharon}, \&
  {Gladders}}]{Wuyts12}
{Wuyts}, E., {Rigby}, J.~R., {Sharon}, K., \& {Gladders}, M.~D. 2012, \apj,
  755, 73

\bibitem[{{Yabe} {et~al.}(2014){Yabe}, {Ohta}, {Iwamuro}, {Akiyama}, {Tamura},
  {Yuma}, {Kimura}, {Takato}, {Moritani}, {Sumiyoshi}, {Maihara}, {Silverman},
  {Dalton}, {Lewis}, {Bonfield}, {Lee}, {Curtis-Lake}, {Macaulay}, \&
  {Clarke}}]{Yabe14}
{Yabe}, K., {Ohta}, K., {Iwamuro}, F., {Akiyama}, M., {Tamura}, N., {Yuma}, S.,
  {Kimura}, M., {Takato}, N., {Moritani}, Y., {Sumiyoshi}, M., {Maihara}, T.,
  {Silverman}, J., {Dalton}, G., {Lewis}, I., {Bonfield}, D., {Lee}, H.,
  {Curtis-Lake}, E., {Macaulay}, E., \& {Clarke}, F. 2014, \mnras, 437, 3647

\bibitem[{{Yates} {et~al.}(2012){Yates}, {Kauffmann}, \& {Guo}}]{Yates12}
{Yates}, R.~M., {Kauffmann}, G., \& {Guo}, Q. 2012, \mnras, 422, 215

\bibitem[{{Yuan} {et~al.}(2013{\natexlab{a}}){Yuan}, {Kewley}, \&
  {Rich}}]{Yuan13b}
{Yuan}, T.-T., {Kewley}, L.~J., \& {Rich}, J. 2013{\natexlab{a}}, \apj, 767,
  106

\bibitem[{{Yuan} {et~al.}(2013{\natexlab{b}}){Yuan}, {Kewley}, \&
  {Richard}}]{Yuan13a}
{Yuan}, T.-T., {Kewley}, L.~J., \& {Richard}, J. 2013{\natexlab{b}}, \apj, 763,
  9

\bibitem[{{Zahid} {et~al.}(2012){Zahid}, {Dima}, {Kewley}, {Erb}, \&
  {Dav{\'e}}}]{Zahid12b}
{Zahid}, H.~J., {Dima}, G.~I., {Kewley}, L.~J., {Erb}, D.~K., \& {Dav{\'e}}, R.
  2012, \apj, 757, 54

\bibitem[{{Zahid} {et~al.}(2013){Zahid}, {Geller}, {Kewley}, {Hwang},
  {Fabricant}, \& {Kurtz}}]{Zahid13}
{Zahid}, H.~J., {Geller}, M.~J., {Kewley}, L.~J., {Hwang}, H.~S., {Fabricant},
  D.~G., \& {Kurtz}, M.~J. 2013, \apjl, 771, L19

\bibitem[{{Zaritsky} {et~al.}(1994){Zaritsky}, {Kennicutt}, \&
  {Huchra}}]{Zaritsky94}
{Zaritsky}, D., {Kennicutt}, Jr., R.~C., \& {Huchra}, J.~P. 1994, \apj, 420, 87

\end{thebibliography}

%-------------------------------------------------------------------------------------------------------------
\begin{appendix}
%-------------------------------------------------------------------------------------------------------------
\section{Integration Resolution} \label{sec:app:res}

To trace star formation histories through time, we must choose a duration for our resolution time step, $\Dt$.  Time steps that are too long may not sufficiently resolve the time-scales for the star formation rate correlation, the passive mode (\S\ref{sec:scatt:passive}), or additional starbursts (\S\ref{sec:app:sb}).  Moreover, large values of $\Dt$ are vulnerable to the scenario where a large fluctuation in the star formation rate implies that a mass much greater than the current stellar mass of the galaxy was produced in the previous time step, a phenomenon that leads to an artificially low value of $\zi$.  We find that $\Dt=10^{7}\,\yr$ strikes a fair compromise between minimizing these effects and maintaining a reasonably small computation time acceptable for producing large statistical samples of runs.

%-------------------------------------------------------------------------------------------------------------
\section{Additional Starbursts} \label{sec:app:sb}

\begin{figure*}
\begin{center}
\includegraphics[width=\textwidth,trim=50 200 20 0,clip]{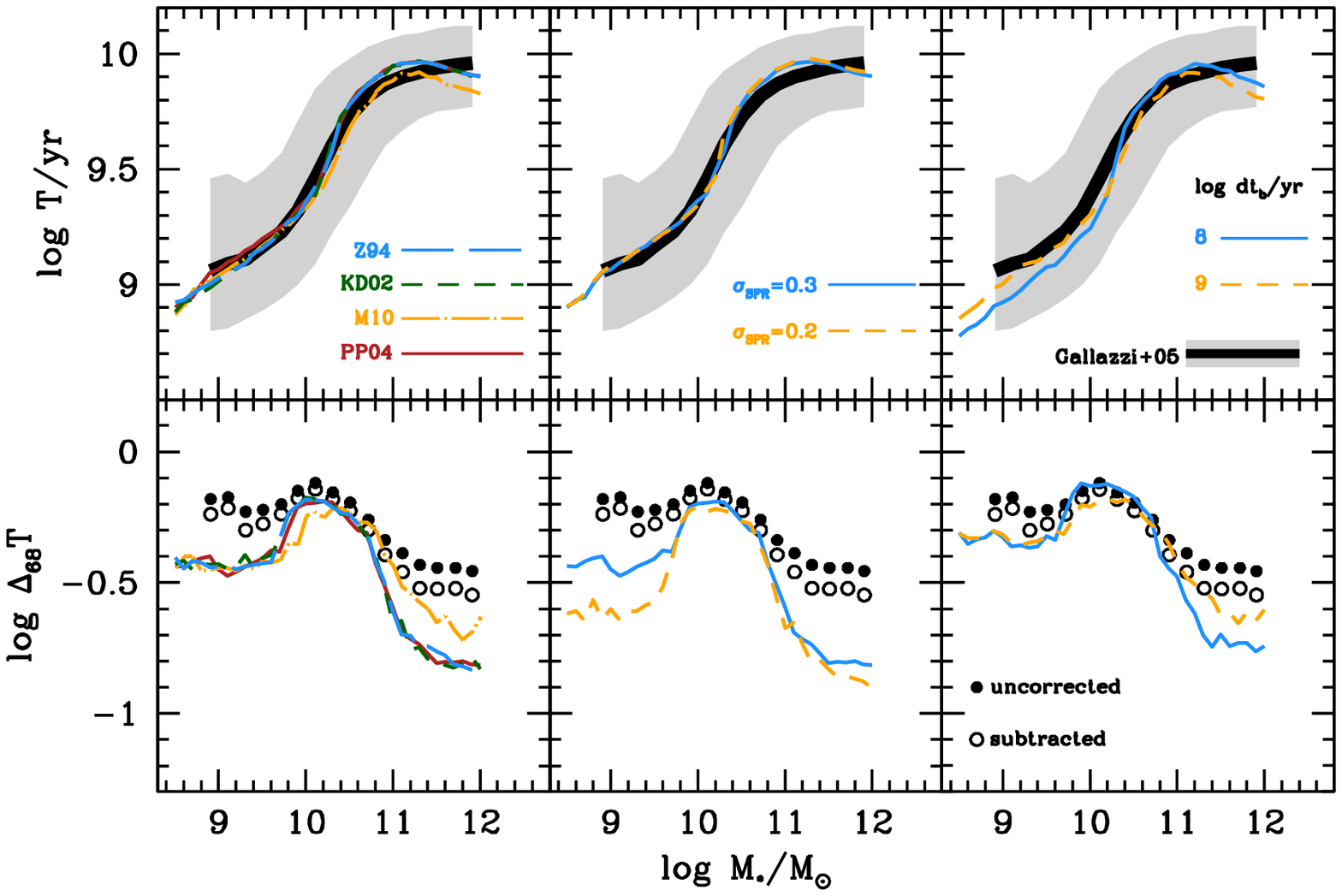}
\caption{\label{fig:scatter2} 
The median (top row) and central 68\% interval range (bottom row) for luminosity-weighted, galaxy-averaged stellar age as a function of total stellar mass at $z=0$.  The presentation of the \citet{Gallazzi05} observations is identical to that in Fig.~\ref{fig:scatter1} with thick lines and shaded regions in the top row indicated median and scatter in ages and filled and open circles in the bottom row showing the uncorrected and subtracted central 68\% interval, respectively.  In the left column, solid (blue), dot-dashed (orange), short-dashed (green), and long-dashed (red) curves compare results using, for the luminosity-weighting of stellar ages within each model galaxy, the FMR calibrations of PP04, M10, KD02, and Z94, respectively.  Passive galaxies are included in the model, as is normal star formation with $\Dtc=10^{9}\,\yr$.  In the middle column, the solid (blue) and dashed (orange) lines show the effect of changing the logarithmic scatter in the star formation rate at fixed stellar mass from our fiducial values of $\sfrsig=0.3$ to $\sfrsig=0.2$; again passive galaxies are included and $\Dtc=10^{9}\,\yr$.  In the right column, results demonstrate the effect of adding additional starbursts $3\times0.3$ dex above the mean star formation rate in 10\% of galaxies, each lasting for $\Dtb=10^{8}$ (solid, blue) or $10^{9}\,\yr$ (orange, dashed), in additional to normal star formation with $\Dtc=10^{9}\,\yr$ and a passive phase.
}
\end{center}
\end{figure*}

In this section, we test the effect of an additional starburst contribution to galaxy star formation histories that complete observations of the star-forming main sequence may not reflect but may, nevertheless, be necessary to explain the density profiles of dwarf systems \citep[e.g.,][]{Weisz12, Amorisco14, Madau14} and observed in a lensed dwarf systems at high redshift \citep{Jones14}.  Such bursts would represent an further random process that produces spikes in the star formation rate, perhaps as some tidal event funnels gas into the central regions of the galaxy.  These types of systems are not specifically excluded from the samples from which \citet{Whitaker12} derived equation~\ref{eq:sfr}, but their contribution was likely small.  Thus, we refer to the model of star formation described in this section as ``additional'' starbursts, with the implication that we may already have included some form of this process in the method described in \S\ref{sec:scatt:norm}.  In effect, these extra bursts increase the scatter in our model star formation histories and the resulting spread in age and stellar metallicity at fixed $\mnow$.  They also tend to minimize the differences between different values of $\Dtc$ as they come to dominate the fluctuations in star formation rate. 

Given the scope of the current study, we focus on obtaining a rough estimate of the potential contribution of such bursts to the scatter in stellar age and metallicity and are rather less interested in developing a physically accurate picture for including them.  For example, we ignore the addition of stellar mass during mergers that may be associated with the bursts and avoid the extra complications of tracing star formation histories for entire merger trees.  Ignoring this branching of the merger tree may over-estimate the scatter in the star formation history since the merging of two independent histories will tend to smooth out fluctuations from either one.  However, this simplification does not significantly bias our results if the stellar mass added through mergers is not the primary mechanism for galaxy growth \citep[e.g.,][]{Leitner12}.  

To describe the additional starbursts in this section, we construct a three-parameter scheme in which a fraction $\fsb$ of star-forming galaxies, rather than having star formation rates drawn from the log-normal distribution appropriate for normal, star-forming galaxies, instead form stars at a rate $\Nb$ standard deviations above the mean---that is, $\Nb=(\log\sfr-\lsfrave)/\sfrsig$---for a period lasting $\Dtb/\Dt$ time steps.  For additional simplicity, we assume that $\fsb$, $\Nb$, and $\Dtb$ are independent of stellar mass and redshift, however, to allow for the possibility that additional starbursts have a distinct physical source from fluctuations in ``normal" star formation, we do not require that $\Dtb=\Dtc$.  Since our goal is to assess whether additional starbursts play a significant role in the scatter in galaxy star formation history rather than to accurately model the relevant physical processes, we further set representative values of $\fsb=0.1$ and $\Nb=3$ and consider two values of burst duration: short bursts with $\Dtb=10^{8}\,\yr$ and long bursts with $\Dtb=10^{9}\,\yr$.  More common or more intense bursts likely would easily have stood out in the \citet{Whitaker12} data and thus already be accounted for in \S\ref{sec:scatt:norm}.  Note that models in which no additional starbursts are included can be thought of as ones with $\fsb=0$.

As in the case for incorporating the passive phase in \S\ref{sec:scatt:passive}, the probability for whether a galaxy begins a burst in a given time step is related to the change in the number of starbursts over the redshift interval.  The procedure is as follows.  In the first time step for which a galaxy is designated as star-forming (either at $z=0$ or after any passive phase), we give the galaxy a probability $\fsb\,(1-\fpass)$ of beginning a starburst.  During a burst, the galactic star formation rate is set to be $\log\sfr=\lsfrave+\Nb\,\sfrsig$ rather than being drawn from the log-normal distribution of \S\ref{sec:scatt:norm}, and each burst lasts for a time $\Dtb$ (i.e., for $\Dtb/\Dt$ time steps).  In all time steps for which a star-forming galaxy is not in a starburst phase, the probability, $\Psb$, of beginning a new starburst in that time step is given by the ratio of the change in the number of starburst galaxies over that time step to the total number of non-bursting, star-forming galaxies:
\begin{equation}\label{eq:Pm}
\Psb(\mstar, z)=\fsb\,\left\{\frac{\dd}{\dd z}\,\ln\left[(1-\fpass)\,\frac{\dd n}{\dd\mstar}\right]\,\dd z+\frac{\Dt}{\Dtb}\right\},
\end{equation}
where the first term in the curly brackets accounts for the changing number of star-forming galaxies and the second accounts for galaxies finishing their starburst phases during the time step.  This procedure ensures that the fraction of a large population of star-forming galaxies in a starburst phase in any time step is always $\fsb$.  As a result, a galaxy that remains star-forming for a time, $T_{\rm SF}$, experiences of order $\sim\fsb\,T_{\rm SF}/\Dtb$ bursts.

For a burst model with values of $\fsb=0.1$, $\Nb=3$ and $\Dtb=10^{8}\,\yr$ interspersed among normal star formation correlated over $\Dtc=10^{9}\,\yr$ and a late passive phase, the bursts create large spikes in the star formation rate, but, ultimately, the buildup of stellar mass is not very different from that with $\fsb=0$.

%-------------------------------------------------------------------------------------------------------------
\section{Sensitivity of Ages to Model Parameters} \label{sec:app:test}

In Figure~\ref{fig:scatter2}, we explore the influence on our model ages of variations in the FMR (left column), differing amounts of scatter in the normal mode of star formation at fixed stellar mass (middle column), and the inclusion of additional star bursts (right column).  The FMR only enters the calculation of stellar ages through the luminosity weighting, and we find that our results are relatively insensitive to the specific choice of calibration for the relation---though results for the M10 relation produce a slightly lower median and increased scatter at the very high-mass end.  Observed uncertainties in $\sfrsig$ are also of only minor importance.  Similarly, while our additional starbursts tend to decrease the median ages at the very low- and high-mass ends of the population below the observed values and somewhat increase the amount of scatter at all masses, they do not alter the results significantly.  Whether or not extra bursts exist as a real component of galaxy formation, the ``additional starbursts" described in \S\ref{sec:app:sb} are, at least, not necessary for understanding the spread in ages as a function of $z=0$ stellar mass.

%-------------------------------------------------------------------------------------------------------------
\section{Supplementary Tables} \label{sec:app:tab}

To facilitate the use of our results by the community, we provide electronic tables of our results in the Supplementary Data.  We generated each table from $10^3$ galaxy realizations in bins of final mass from $10^{8.5}$--$10^{12}\,\msun$.  The column headings, from left to right, are (1) stellar mass; (2--4) 16th, 50th, and 84th percentile of luminosity-weighted age; (5--7) 16th, 50th, and 84th percentile of luminosity-weighted stellar metallicity; (8--10) 16th, 50th, and 84th percentile of mass-weighted age; (11--13) 16th, 50th, and 84th percentile of mass-weighted stellar metallicity; (14--16) 16th, 50th, and 84th percentile of the redshift at which 90\% of the final stellar mass is reached; (17--19) 16th, 50th, and 84th percentile of the redshift at which 50\% of the final stellar mass is reached (i.e., $\zhalf$); and (20--22) 16th, 50th, and 84th percentile of the quenching redshift (see Table~\ref{tab:sample} for a partial example).  Values of ``-999" for the quenching redshift indicate that a value could not be determined because not enough realizations resulted in a passive system at the final redshift.  We include tables with results for star-forming only, passive only, or all galaxies with $\zf=0$, 0.1, 0.25, 0.5, 1.0, 1.5, and 2.0 assuming $\Dtc=10^{9}$ and using the PP04 FMR calibration.  Additionally, we provide results using the KD02, Z94, and M10 FMR calibrations at $\zf=0$.  

\end{appendix}
%-------------------------------------------------------------------------------------------------------------
\clearpage
\begin{landscape}

\begin{table}
\begin{flushleft}
\caption{Sample Table}
\begin{tabular}{cccccccccccccccc}
\hline
$\mfinal$ & $T^{\rm LW}_{16}$ & $T^{\rm LW}_{50}$ & $T^{\rm LW}_{84}$ & $Z^{\rm LW}_{16}$ & $Z^{\rm LW}_{50}$ & $Z^{\rm LW}_{84}$& $T^{\rm MW}_{16}$ & $T^{\rm MW}_{50}$ & $T^{\rm MW}_{84}$ & $Z^{\rm MW}_{16}$ & $Z^{\rm MW}_{50}$ & $Z^{\rm MW}_{84}$ & . & . & . \\
\hline
8.5 & 8.69714 & 8.89606 & 9.06296 & -0.990586 & -0.965648 & -0.947322 & 9.28166 & 9.39568 & 9.49006 & -1.08692 & -1.07315 & -1.06288 & . & . & .\\
8.6 & 8.72299 & 8.91429 & 9.09394 & -0.962493 & -0.93594 & -0.918917 & 9.30402 & 9.42289 & 9.5099 & -1.0587 & -1.04433 & -1.03491 & . & . & .\\
8.7 & 8.74968 & 8.934 & 9.11533 & -0.933756 & -0.906849 & -0.89003 & 9.32176 & 9.43818 & 9.53025 & -1.03102 & -1.01704 & -1.00736 & . & . & .\\
8.8 & 8.76695 & 8.97311 & 9.14073 & -0.903026 & -0.87624 & -0.860836 & 9.35914 & 9.46609 & 9.54669 & -1.0027 & -0.988028 & -0.978879 & . & . & .\\
8.9 & 8.78893 & 9.00361 & 9.17176 & -0.872773 & -0.847879 & -0.83193 & 9.37691 & 9.48733 & 9.5702 & -0.973877 & -0.959993 & -0.951397 & . & . & .\\
9 & 8.8373 & 9.02734 & 9.18924 & -0.843413 & -0.817594 & -0.804073 & 9.41356 & 9.50957 & 9.58702 & -0.943983 & -0.931822 & -0.923149 & . & . & .\\
. & . & . & . & . & . & . & . & . & . & . & . & . & & &\\
. & . & . & . & . & . & . & . & . & . & . & . & . & & &\\
. & . & . & . & . & . & . & . & . & . & . & . & . & & &\\
\hline
\end{tabular}\label{tab:sample}
\end{flushleft}
\end{table}

\begin{table}
\begin{flushright}
%\caption{Sample Table (cont.)}
\begin{tabular}{cccccccccccc}
\hline
. & . & . & $z_{\rm 9/10,\,16}$ & $z_{\rm 9/10,\,50}$ & $z_{\rm 9/10,\,84}$ & $z_{\rm 1/2,\,16}$ & $z_{\rm 1/2,\,50}$ & $z_{\rm 1/2,\,84}$ & $z_{\rm q,\,16}$ & $z_{\rm q,\,50}$ & $z_{\rm q,\,84}$ \\
\hline
. & . & . & 0.0262418 & 0.0590655 & 0.101895 & 0.151021 & 0.229398 & 0.31799 & -999 & -999 & -999\\
. & . & . & 0.0285369 & 0.0614686 & 0.11043 & 0.155571 & 0.248789 & 0.335151 & -999 & -999 & -999\\
. & . & . & 0.0316087 & 0.066299 & 0.11733 & 0.170322 & 0.267627 & 0.370653 & -999 & -999 & -999\\
. & . & . & 0.0331495 & 0.0785186 & 0.127803 & 0.193973 & 0.283689 & 0.389027 & -999 & -999 & -999\\
. & . & . & 0.0354671 & 0.0859507 & 0.145598 & 0.204621 & 0.30231 & 0.414212 & -999 & -999 & -999\\
. & . & . & 0.0409042 & 0.0901129 & 0.154659 & 0.225379 & 0.327095 & 0.432331 & -999 & -999 & -999\\
& & & . & . & . & . & . & . & . & . & .\\
& & & . & . & . & . & . & . & . & . & .\\
& & & . & . & . & . & . & . & . & . & .\\
\hline
\end{tabular}
\end{flushright}
All galaxies with $\zf=0$ and assuming $\Dtc=10^{9}\,\yr$ and the PP04 FMR calibration.\\
Col. (1): Stellar mass at $\zf$ in log units of solar masses.  Col. (2--4): 16th, 50th, and 84th percentile of luminosity-weighted age ($T^{\rm LW}$) in log units of years.  Col. (5--7): 16th, 50th, and 84th percentile of luminosity-weighted stellar metallicity ($Z^{\rm LW}$) in log units of solar metallicity.  Col. (8--10): 16th, 50th, and 84th percentile of mass-weighted age ($T^{\rm MW}$) in log units of years.  Col. (11--13): 16th, 50th, and 84th percentile of mass-weighted stellar metallicity ($Z^{\rm MW}$) in log units of solar metallicity.  Col. (14--16): 16th, 50th, and 84th percentile of the redshift at which 90\% of the final stellar mass is reached ($z_{\rm 9/10}$).  Col. (17--19): 16th, 50th, and 84th percentile of the redshift at which 50\% of the final stellar mass is reached ($\zhalf$).  Col. (20--22): 16th, 50th, and 84th percentile of the quenching redshift ($\zq$).  Full tables are available online.
\end{table}

\end{landscape}
\clearpage
%-------------------------------------------------------------------------------------------------------------

\end{document}